# Sportoonizer: Augmenting Sports Highlights' Narration and Visual Impact via Automatic Manga B-Roll Generation

SIYING HU, City University of Hong Kong, China
XIANGZHE YUAN, City University of Hong Kong, China
JIAJUN WANG, City University of Hong Kong, China
PIAOHONG WANG, City University of Hong Kong, China
JIAN MA, City University of Hong Kong, China
ZHIYANG WU, City University of Hong Kong, China
QIAN WAN, City University of Hong Kong, China
ZHICONG LU, City University of Hong Kong, China

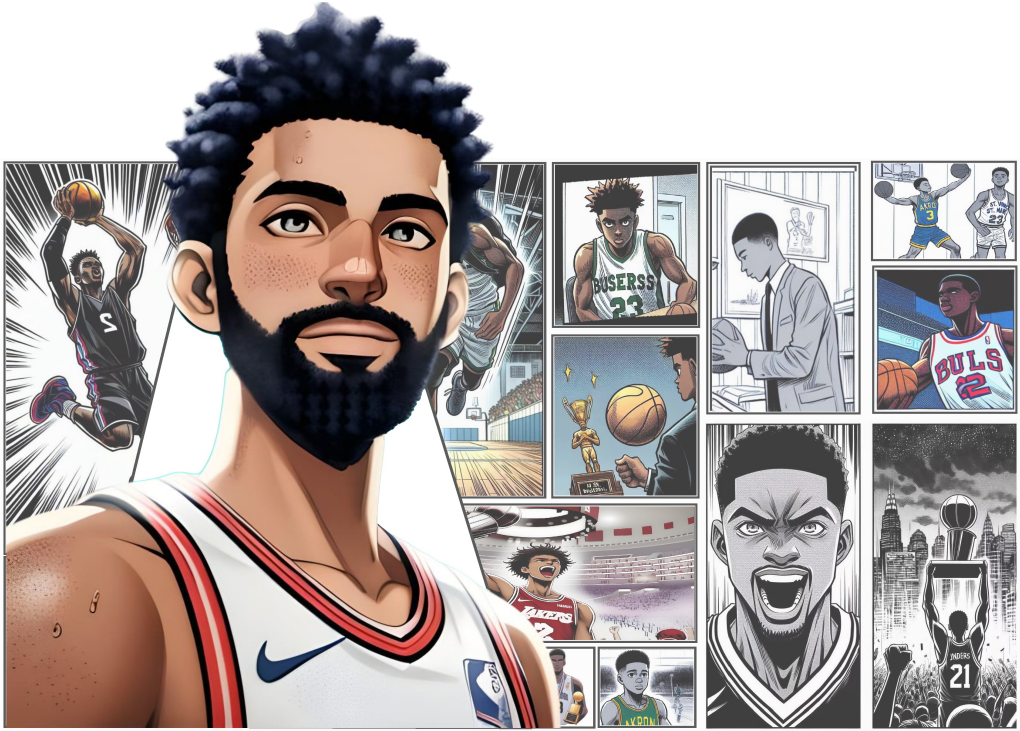

Fig. 1. *Sportoonizer* augmenting sport highlights' narration and visual impact through automatic Manga B-Roll generation. It seamlessly merges dynamic manga sequences with live-action footage, enriching the visual tapestry and deepening the narrative scope of the content.

Sports highlights are becoming increasingly popular on video-sharing platforms. Yet, crafting sport highlight videos is challenging, which requires producing engaging narratives from different angles, and conforming to different platform affordances with constantly changing audiences. Many content creators therefore create derivative work of the original sports video through manga styles to enhance its expressiveness. But manually creating and inserting tailored manga-style content can still be time-consuming. In this paper, we introduce *Sportoonizer*, a system that embeds the pipeline for the automatic generation of manga-style animation for highlights in sports videos and insertion into original videos. It seamlessly merges dynamic manga sequences with live-action footage, enriching the visual tapestry and deepening the narrative scope of the content. By leveraging social media data and advanced generative AI models, *Sportoonizer* crafts compelling storylines that encapsulate the intensity of sports moments and the personal journeys of athletes. Our evaluation study demonstrates that integrating manga B-rolls significantly enhances viewer engagement, evoking greater visual interest, emotional connection, and empathetic response towards the athletes' stories in the viewing experience.

CCS Concepts: • **Human-centered computing** → **Interactive systems and tools**.

Additional Key Words and Phrases: Sports video, Video Augmentation, Manga, Comic, Engagement, Narrative, Storytelling

# 1 INTRODUCTION

Sports highlights have become increasingly popular on video-sharing platforms, providing content creators a unique opportunity to engage with their audience across differnt backgroud. With the emergence of video-sharing platforms such as YouTube, Bilibili[1], and TikTok, fans can access exciting moments from sports events quickly and conveniently [77]. These platforms have become the go-to destinations for viewers to access thrilling moments from sports events, express their interests, and participate in discussions [18]. The availability of captivating video content has enabled remote audiences to experience the attraction of sports events, as they can now view exciting moments captured by photographers and broadcasting entities through video-sharing platforms [42, 58].

Many content creators adapt derivative work to craft captivating content that captures the attention of the public [44] with Generative AI models. This involves taking existing content and modifying it to create something new and engaging. One such approach is the use of *motion graphics*, which combines live-action and animation to produce visually stunning videos. However, creating crafting sport highlight videos is challenging, which requires producing engaging narratives from different angles, and conforming to different platform affordances with constantly changing audiences. This process also requires a solid background in sports knowledge. Many content creators therefore create derivative work of the original sports video through manga styles to enhance its expressiveness. But manually creating and inserting tailored manga-style content can still be time-consuming. Prior work has proposed automated solutions with embedded visualizations overlaying sports videos for different contexts such as knowledge sharing [12], accessibility, content marketing [58], and customized professionals' needs (e.g., sports analysis [49] and motion trajectories [64]). These solutions aim to enhance the viewing experience and engage viewers without distracting them from the main action [12, 58].

---

[1]Bilibili: https://www.bilibili.com/

Transmedia storytelling[2], a concept introduced by Henry Jenkins, refers to the technique of telling a single story or story experience across multiple platforms and formats using current digital technologies [33]. Transmedia narrative has been explored in various contexts, including sports journalism [2, 14] and social media narratives [57, 67, 70]. This approach offers unique opportunities for enhancing narrative depth and audience engagement. Concurrently, research has highlighted the potential of visual elements in augmenting storytelling impact. Studies have examined the use of graphic elements in educational videos [56, 63] and the power of visual storytelling in news media [28, 29], demonstrating how visuals can significantly enhance narrative comprehension and emotional resonance. Of particular interest is the potential of manga-inspired visuals in sports broadcasting and marketing [4, 7, 37]. This distinctive visual style not only captures audience attention but also offers innovative ways to augment traditional narratives, potentially increasing both comprehension and emotional impact. Within this context, we introduce *Manga B-roll* as a novel transmedia element enhancing visual narration in sports storytelling [38, 39, 65]. *Manga B-roll* integrates the expressive visual language of manga comics with traditional video B-roll techniques, creating a unique hybrid that enriches narrative depth and visual impact. Our approach attempted to introduce *Manga B-roll* as an interactive medium that integrates manga-style visual elements with traditional sports footage and enhances narrative depth in sports highlights. This method aims to visualize tactical concepts, represent game dynamics, and convey emotional aspects that are often implicit in conventional sports coverage. Specifically, this research include the participatory design in which we explored the needs and design considerations of using *Manga B-rolls* for sports telling, and developed the proof-of-concept system for evaluation.

To achieve our goals, we introduce *Sportoonizer*, an LLM-powered system designed to incorporate manga-style B-rolls into sports videos. In this project, our system's automated pipeline seamlessly blends dynamic manga sequences with live-action footage, creating a visually intricate tapestry that expands the narrative scope of the content. By leveraging social media data (e.g., athlete wiki text recording and public graph) and advanced generative AI models, *Sportoonizer* creates compelling storylines that capture the intensity of sports moments and the personal journeys of athletes. The pipeline for the input video consists of four distinct stages: the identification and extraction of sports events, the generation of *Manga B-roll*, the segmentation and embedding of frames, and the rendering of video. It facilitates content creation by leveraging dynamic storytelling elements of manga for motion highlights.

We demonstrate the efficacy of our automated pipeline by presenting a wide variety of auto-generated example videos of popular sports. Our evaluation study shows that incorporating *Manga B-rolls* significantly boosts viewer engagement. It generates greater visual interest, emotional connection, resonance, and empathetic response towards the athletes' stories in the viewing experience. *Sportoonizer* provides a unique and innovative solution to the challenges of crafting short-form videos that resonate with an ever-changing audience. With its ability to create visually stunning and emotionally engaging content, *Sportoonizer* has the potential to revolutionize the way sports highlight videos are created and consumed. In summary, our work makes the following contributions:

---

[2]Transmedia storytelling: https://en.wikipedia.org/wiki/Transmedia_storytelling.

- We conduct a formative study of participatory design of existing sports highlights videos and *Manga B-roll* examples to identify how the *Manga B-roll* can be designed and used for improving narration and visual compelling.
- We show how this *Manga B-roll* can be automatically generated from an input video using transformer model leveraging a combination of computer vision and generative AI models.
- We design and implementation of the *Sportoonizer* and show how the input video can then be rendered into a transmedia video that deepening sports narratives and fostering viewer engagement.
- We conduct two-part user evaluation to understanding of how *Manga B-roll* afforded augmenting narration and visual impact in sports highlights.

## 2 RELATED WORK

*Sportoonizer* builds on prior research on supporting viewers' experience and engagement with sports videos, *Manga B-roll* and Transmedia Storytelling, and AI-supported content creation.

### 2.1 *Manga B-roll* and Transmedia Storytelling

*Manga B-roll* is an innovative transmedia element that enriches sports storytelling by fusing manga-style visuals with traditional video content [24, 62, 76]. The manga style is an ideal choice for enhancing the narrative impact of sports videos due to its expressive visual language, broad cultural appeal, and strong emotional resonance. By incorporating manga-style B-roll into sports videos, we can present a more diverse range of narrative content and more effectively convey the emotions and drama inherent in sports events. Additionally, the graphical elements of manga provide a more intuitive way to communicate complex tactical and dynamic information, thereby improving viewer comprehension and engagement. This technique builds upon transmedia storytelling, which distributes narrative elements across multiple platforms to create a cohesive entertainment experience [19, 59, 69]. For example, in the film "Scott Pilgrim vs. the World" (2010), comic elements were extensively used for scene transitions and visual effects, blending live-action with comic book aesthetics to create a unique visual language that enhanced the original comic's style and provided an immersive viewing experience [86]. Similarly, the animated film "Spider-Man: Into the Spider-Verse" (2018) fully integrated comic book aesthetics, including panels and speech bubbles, to create a distinctive visual style that not only faithfully reproduced comic visuals but also effectively portrayed the concept of a multiverse [55]. In sports media, *Manga B-roll* supplements conventional footage by visualizing tactical [45, 85], emotional [81], story construction [20] and contextual aspects of highlights [34]. Unlike standard sports broadcasting techniques [25, 50], *Manga B-roll* employs manga-inspired illustrations to convey both explicit and implicit information [22, 41]. It applies artistic modifications to capture the essence of sports events, creating a medium that offers rich semantics and evokes viewer emotions [72]. Media researchers have explored various visual enhancements in sports coverage [27, 78], demonstrating the effectiveness of graphic elements in conveying complex information, such as content marketing [31, 71] and sense-making [9, 36]. However, the potential of manga-style visuals in sports broadcasting remains largely unexplored. Studies suggest that viewer experience with such visual elements depends on their integration and presentation [16], with concerns about potential impacts on perceived authenticity [54].

Our research addresses this gap by investigating *Manga B-roll* as a storytelling tool in sports highlights. We explore its design considerations and potential applications, aiming to enhance sports narratives and viewer engagement. Through a study using a proof-of-concept system, we demonstrate how *Manga B-roll* can potentially transform sports storytelling in the digital age.

## 2.2 Supporting Viewer's Experience and Engagement with Sports Videos

In order to enhance the audience experience and engagement of sports videos, existing work mainly includes exploring the impacts of audience experience and innovative ways of content creation.

*2.2.1 Effects of Viewer Experience.* There are various factors that can impact a viewer's experience when watching sports videos. Many studies have explored the significance of visual enjoyment [5], emotional engagement [47], and interactivity in sports video [35, 87]. For example, through video quality optimization [89], multi-angle shooting [8], visualization [12, 49], and virtual/augmented reality [83], viewers can now enjoy a more realistic and immersive experience. These technologies improve the clarity and detail of sports content, resulting in increased viewer satisfaction. However, more than just the technology has improved the viewer's experience. Interactive storytelling also plays a critical role in eliciting emotional resonance from viewers and has become a rich domain to be explored [39]. In SCoReS, the system automatically suggests stories about athletes' struggles, victories, and setbacks during games, establishing a strong emotional connection with the audience [40]. In addition, the emergence of social media platforms and interactive live-streaming features have transformed how viewers engage with sports videos [35, 87]. Real-time discussions, polls, and direct interactions with athletes and commentators have turned passive viewing into an active, community-building experience. The current research contributes to this line of research by proposing a novel form of engagement with sports videos that incorporates *Manga B-rolls* for enhanced narratives and visual impact.

*2.2.2 Content Creation in Sports Videos.* In order to create captivating sports video content, it's vital to incorporate innovative techniques that can engage viewers. These techniques can include using diverse subject matter [60], multiple camera perspectives [15], and personalized recommendations [66]. The sports industry has recently seen a revolution with the introduction of unique elements such as first-person and 360-degree views of athletes, as well as drone shooting [53]. These state-of-the-art technologies offer viewers a fresh perspective that enhances their understanding and appreciation of sports movements, elevating the overall viewing experience. To cater to the diverse interests of the audience, sports programs offer a multi-dimensional view of sports and athletes, including live broadcasts, behind-the-scenes footage, athlete interviews, and sports highlights [21]. By diversifying content, the appeal of sports videos can be expanded, and the audience can be engaged in a profound way. Besides, there are some sports video platforms that leverage audience data to artificial intelligence to provide personalized recommendations and viewing experiences based on individual preferences [77]. By analyzing the audience's viewing habits and preferences, AI can suggest relevant content that enhances the audience's satisfaction and creates a customized viewing experience.

## 2.3 AI-supported Media Content Creation

AI-supported media content creation is transforming the way content is produced by leveraging LLMs, computer vision, and machine learning [13, 88]. Systems like LAVE [82] and ExpressEdit[79] enable video editing through natural language and sketching, streamlining the editing process and making it more accessible. Other innovations, such as automated music video lyric synchronization [52] and sound effect matching [48], enhance the audio-visual experience. Xia et al., developed Crosscast [88], to enhance the experience of audio travel podcasts by incorporating visual elements. These AI-driven tools democratize content creation, allowing for more engaging and immersive multimedia narratives. Generative AI (GenAI) has made significant progress in media content creation, particularly in automatic video creation and manga filters on social media platforms. Automatic video creation utilizes AI algorithms to efficiently generate high-quality videos based on specific inputs or parameters. Lyu et al. [51] found that GenAI has been widely used in marketing, art, entertainment, programming, and education to create digital art, assist in marketing strategies, and provide innovative tools for content creators. On video-sharing platforms, one of the popular applications is manga filters, transforming photos and videos into manga-style artworks through the use of various artistic techniques [32]. These advancements in GenAI have revolutionized content creation and provided new avenues for creativity and engagement in the digital landscape. However, our work distinguishes itself by focusing on leveraging generative AI to augment sports videos with generated *Manga B-rolls* that align with existing video content, providing viewers with rich narratives and visual experiences. This approach goes beyond the widely used manga filters, offering a more dynamic and engaging form of content creation.

## 3  DESIGN CONSIDERATIONS OF SYSTEM AND *MANGA B-ROLL*

To explore the feasibility of automatically generating *Manga B-rolls* for augmenting sports highlights, we conducted a participatory design workshop with sports stakeholders. The workshop drew inspiration from the visual language of *Manga B-rolls* to inform the design of a system for enhancing the narration and visual appeal of sports highlights.

### 3.1  Method: Participatory Design Workshop

To further explore the user practices, requirements and suitable design factors for the narrative design through *Manga B-roll*, we conducted participatory design workshop with 15 participants (can see demographic in subsection 11.1) inspired by Zytko et. al., work [90]. These participants are all stakeholders in the sports sector, including those from the sports industry (sports enthusiasts and spectators, athletes, coaches, students, teachers, sports commentators), the video production industry (video content creators, screenwriters, editors, designers, and relevant technical personnel), and anime enthusiasts. Among them, 1) stakeholders from the sports industry, as direct participants in sports video content, are capable of providing feedback on user preferences and user experiences based on their own experiences. 2) Stakeholders in the video production industry possess professional experience in capturing and editing sports event videos. They understand the production process and objectives, and know how to attract more audiences through innovative storytelling methods. They can provide content guidance and narrative design suggestions for our system design. And 3) Anime enthusiasts have extensive experience in browsing manga, which can aid in transforming sports content into a manga-style narrative.

We have posted a recruitment notice on social media, where all participants are familiar with sports reporting and video content, and are occasional or regular viewers of sports short videos, receiving compensation according to local standards. We require participants to create B-roll by drawing activities or utilizing provided paper materials based on the content of sports highlights, and to organize the overall video into a story they believe will be appealing. These design resources include a variety of sports highlights videos provided, sports backgrounds from the videos, sports knowledge and relevant athlete profiles, storyboards, sticky notes, and the sports manga, comic transition examples, and drawing tools we have supplied. We asked participants to design a story that prominently featured sportsmanship. Subsequently, we conducted semi-structured interviews to understand participants' reflections on their narrative designs, their preferences in expression, and the considerations in their design process.

Based on this design outputs analysis and the literature on comic visual communication [1, 17, 73], cinematographic guidelines about B-roll insertion [30, 61, 68], transmedia narrative strategies in sports journalism [16, 65], we derived the final [**D**]esign [**C**]onsiderations for *Sportoonizer*. Our analysis also indicated [**T**]enants of *Manga B-roll* for how these content can enhance the narrative depth and visual appeal of sports highlights through various means, including diverse content expressions, insertion timing, and timeline positioning. This process was done by three researchers, and after several rounds of discussion and iteration, they reached an agreement.

## 3.2 Design Considerations of System

Based on the results of participatory design, we propose the following design considerations for narratives emerged and implement a system to support sports highlights' narration augmenting and video impact improvement through *Manga B-roll*.

*3.2.1* **[DC1] Generating Manga B-roll in suitable timeline to support storytelling narration**. Most participants indicated that for *Manga B-roll* to maximize its impact on video storytelling, it needs to be generated at the right moment. They suggested that the system should be able to produce content-related *Manga B-roll* based on the original video input and insert it at the appropriate timing to enhance narrative effectiveness. Ideally, *Manga B-roll* should be able to automatically identify where narrative threads are missing in the input video, generate supplementary narrative content, and accurately insert it for saving time.

*3.2.2* **[DC2] Maintaining coherence and readability of transmedia narratives for ease of understanding.** All participants agreed that maintaining the coherence and readability of the overall content is crucial in the process of transmedia narratives stitching, which also impacts the audience's experience and comprehension when viewing sports highlights. In terms of visuals, they recommend employing comic transition effects such as "action-to-action," "subject-to-subject," and "scene-to-scene" to achieve seamless stitching and visual transitions. Regarding content, they suggest continuing the original storyline based on the understanding of the input video to create new narrative content that is both connected and developmental.

*3.2.3* **[DC3] Enabling flexible adjustment of narrative center of gravity and sequence.** Some participants suggested that the system should support flexible adjustments to the narrative order (follow or deviate from linearity) of the original video input (the A-roll) and *Manga B-roll* to highlight different emphases and expressions method to help smooth storytelling. Taking advantage of the generative AI models capabilities, users should be able to intuitively handle and craft compelling narrations and visual effects with the *Manga B-roll* generated by themselves.

## 3.3 Tenants of *Manga B-roll*

By integrating manga-style visual elements with traditional sports highlights footage, *Manga B-roll* functions not only as a visual enhancement tool but also as a significant means of deepening sports narratives and fostering viewer engagement. *Manga B-roll* acts as a visual catalyst that adds an interactive narrative layer to sports highlights. It translates intangible elements such as player close-ups, emotional intensities, and game dynamics into tangible visual expressions, subtly guiding viewer comprehension and emotional connection to the content. *Manga B-roll* can also function as a visual metaphor, playing a pivotal role in the representation and conveyance of complex sports scenarios. Such visual elements can be designed to align with the dynamic nature of sports events by employing commonly observed manga storytelling techniques.

Through analyzing participants' design outputs, we identified all design patterns (can see list in Table 1) about the use cases of *Manga B-roll* and how these content help for sports highlights narration and visual impact. In this work, we selected three most frequently proposed concepts: **[T1] Freeze-frame Moments**, **[T2] Athletic Career Showcase**, and **[T3] Contextual Information** and employed these three styles as *Manga B-roll* implications for narratives emerged. These elements emerged as the most popular design patterns among workshop participants, indicating their potential effectiveness in enhancing sports highlights through *Manga B-roll*. Each style serves a unique purpose when augmenting sports highlights, enhancing both narration and visual impact through automatic generation.

| Frequency | Designed Manga B-roll | Description |
|---|---|---|
| # 13 | Freeze-frame Moments | It captures pivotal instances in sports events by pausing the video and inserting manga-style static images. This style highlights key actions, expressions, or details, enhancing viewer perception and appreciation of crucial moments in the game. |
| # 13 | Athletic Career Showcase | It presents athletes' backgrounds and career highlights using manga-style illustrations. By depicting an athlete's growth, achievements, and key milestones, this method builds an emotional connection between viewers and athletes, enriching the overall viewing experience. |
| # 11 | Contextual Information | It employs manga-style graphics to provide additional background data, stakeholders reaction (e.g., teammates, coaches, competitors, spectators), and historical context. This approach deepens viewer understanding of game situations and crowd reactions, offering atmosphere that enhance engagement with the sport. |
| # 9 | Key Action Analysis | It provides detailed breakdowns of crucial sports movements. This approach uses manga-style illustrations to deconstruct complex actions, highlighting technique, strategy, and biomechanics. By offering frame-by-frame visual explanations, it enhances viewer understanding of skillful plays and critical game moments. |
| # 5 | Friendship Bonds | It showcases the deep connections between teammates. Through manga-style vignettes, it illustrates moments of camaraderie, support, and shared experiences both on and off the field. This method adds emotional depth to the narrative, helping viewers connect with the human side of athletes. |
| # 2 | Rivalry Storylines | It explores the history and dynamics between long-term competitors. Using manga-style flashbacks and character-focused panels, it narrates the evolution of rivalries, key confrontations, and mutual respect between opponents. This method adds dramatic tension and historical context to current matchups. |

Table 1. Design Patterns of Manga B-roll in Participatory Design Workshop. We analyzed the role, appearance time, and content of each design's Manga B-roll in sports highlights narratives and visual enhancements. These have been ranked in descending order based on their frequency of appearance.

## 4 AUTOMATED PIPELINE

The core automated pipeline of *Sportoonizer* consists of four main steps, each designed to address the design considerations and implement the tenants of *Manga B-roll*. This process integrates computer vision algorithms and generative AI models to deeply understand the original video content, identify key moments, construct complete narrative structures, and generate corresponding *Manga B-roll* to enhance the overall viewing experience. Please see the workflow in Figure 2 and example with input video in Figure 4.

### 4.1 Step 1. Understanding Video Content

In the initial stage of the process, the system employs advanced AI models to comprehensively analyze the input sports video. The Blip model [43] processes randomly sampled video frames, generating detailed textual descriptions. These descriptions are then fed into the GPT-4o model for integration and deep understanding, forming a comprehensive cognition of the video content. This process not only identifies specific events and characters in the video but also understands the overall context and potential narrative themes, laying a solid foundation for the subsequent generation of *Manga B-roll*.

### 4.2 Step 2. Identifying Sports Highlights, Sentiment and Narrative Elements

Building upon the understanding of video content, the system proceeds to identify key moments and narrative elements follow the experiment conducted by Gao et.al..[23]. This process is divided into two main parts: first, using the pretrained I3D-NL model(Inflated 3D Non-Local network)[84] to precisely identify sports highlight moments in the video; second, analyzing the previously generated content descriptions through natural language processing techniques to identify key elements in the narrative structure, such as openings, conflicts, climaxes, and conclusions. The I3D-NL model, which combines 3D ConvNets with non-local neural networks, is particularly effective in capturing both spatial and temporal dependencies in video data. Simultaneously, the system marks potential narrative gaps, areas lacking sufficient context or emotional depth. This step provides the necessary materials and framework for constructing a complete narrative storyline. The non-local operation defines a generic non-local operation in deep neural networks as:

$$y_i = \frac{1}{C(x)} \sum_{\forall j} f(x_i, x_j) g(x_j) \tag{1}$$

Here, i is the index of an output position, and j is the index that enumerates all possible positions. The response is normalized by a factor C(x). However, due to the lack of an appropriate sports video highlight dataset for training, we compiled a new dataset specifically for this purpose. Following a similar annotation method to that of Su et al.[74], we labeled 75 videos across seven popular sports with clearly defined highlights, including basketball, soccer, and diving. This dataset was used to train the highlight recognition model employed in our pipeline.

### 4.3 Step 3. Constructing the Narrative Storyline

With a deep understanding of the video content and identification of key elements, the system now proceeds to construct a complete narrative storyline. This process centers around the identified sports highlights while considering the narrative gaps discovered in the previous analysis. The system utilizes the GPT-4o model to generate suggestions for narrative elements to fill these gaps, which may include background information, character introductions, or emotional build-up. The use of GPT-4o allows for contextually relevant and coherent narrative expansions. Ultimately, the system creates a comprehensive narrative outline that includes both the essence of the original video and parts that need to be supplemented through *Manga B-roll*, ensuring the completeness and appeal of the story.

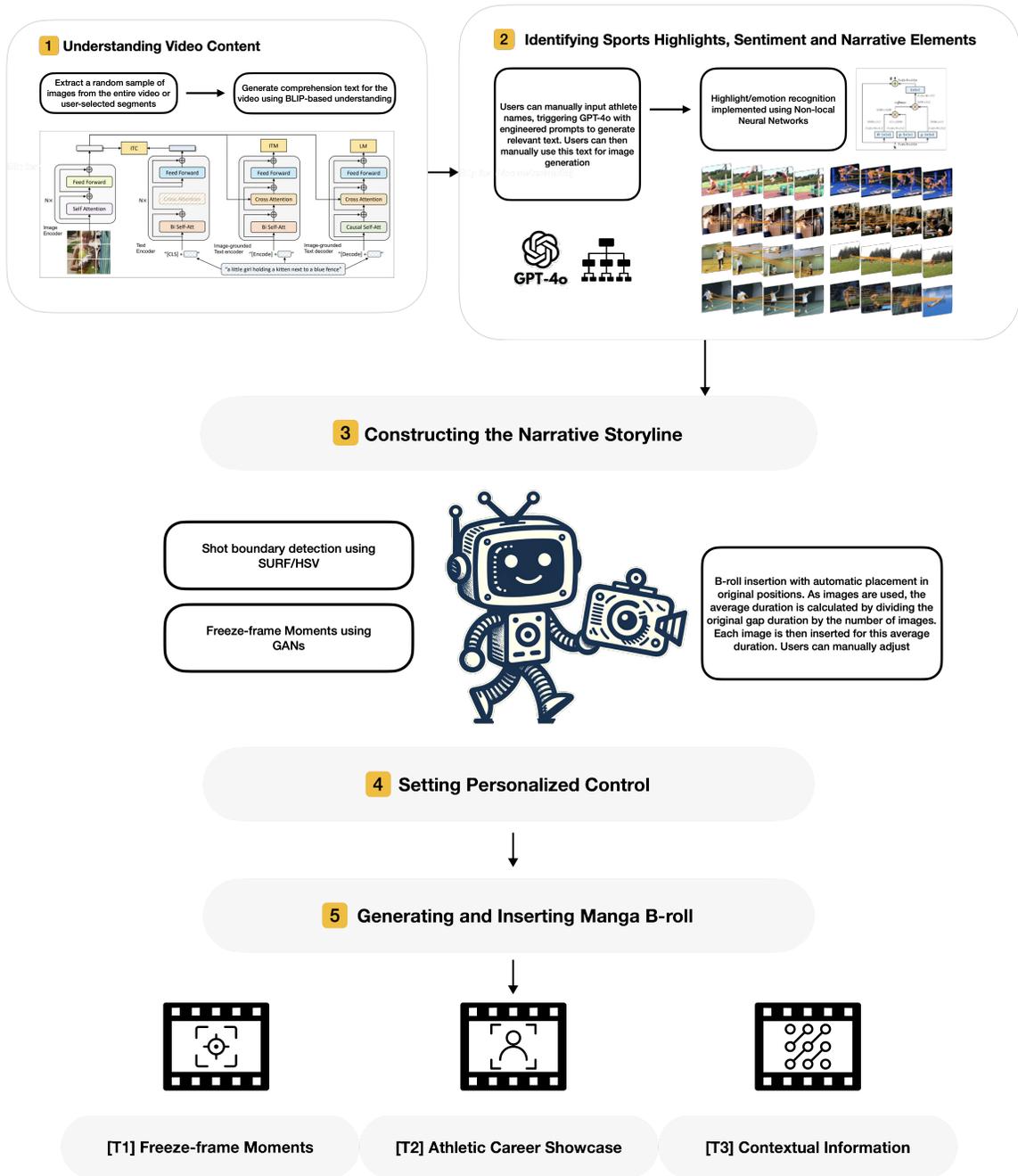

Fig. 2. Auto Pipeline include Five Steps. This process integrates computer vision algorithms and generative AI models to deeply understand the original video content, identify key moments, construct complete narrative structures, and generate corresponding *Manga B-roll* to enhance the overall viewing experience.

## 4.4 Step 4. Setting Personalized Control

To ensure that the final outputs can meet the needs and preferences of different users, the system introduces personalized control options at this stage. Users can select or adjust specific narrative elements, filter style, and density in all timelines, decide which narrative gaps need to be filled through *Manga B-roll*, and adjust the order and emphasis of the narrative. This flexibility allows each user to create a unique visual story according to their preferences, greatly enhancing the system's interactivity and user engagement.

## 4.5 Step 5. Generating and Inserting *Manga B-Roll*

After completing the narrative structure and personalized adjustments, the system generates and inserts the *Manga B-roll*. This multi-faceted process involves several key components to create a comprehensive and engaging visual narrative.

*4.5.1 Sports Highlights Detection.* The system employs advanced computer vision algorithms and machine learning models, particularly the I3D-NL model, to identify and extract key sports highlights from the original video. This process involves analyzing motion patterns, player actions, and game events to pinpoint the most significant and exciting moments forming the narrative's core. The I3D-NL model's ability to capture long-range dependencies in video data makes it particularly suitable for this task.

*4.5.2 Shot Boundaries and Sentiments Analysis.* We employ a combination of SURF (Speeded-Up Robust Features)[3] and HSV color histogram[75] differences for Shot Boundary Detection. This method segments the video into independent shots. From each segment, we extract 64 frames at 10-frame intervals, resizing them to 224×224 pixels for I3D-NL network input.

We continue to use the I3D-NL model for video sentiment analysis. We collected sports videos across seven different types of sports and labeled them with four emotions that commonly appear in sports: excitement, tension, disappointment, and anger. The trained I3D-NL model is then integrated into *Sportoonizer* for sentiment analysis. This analysis helps us understand the emotional context and maintain coherence when inserting *Manga B-roll* content.

*4.5.3 [T1] Freeze-frame Moments.* Building on the detected sports highlights, the pipeline generates manga-style freeze-frame moments. These stylized images capture the essence of key actions or emotions, transforming them into visually striking manga-inspired illustrations that punctuate the narrative and emphasize crucial moments in the sports event. The pipeline employs image-to-image methods that utilize GANs (Generative Adversarial Networks)[26] trained on manga-style artwork (can see example in Figure 3).

*4.5.4 [T2] Athletic Career Showcase.* To provide context and depth to the narrative base on subsection 3.1, the pipeline creates a manga-style showcase of the featured athletes' careers. This involves generating a series of illustrations depicting significant milestones, achievements, or defining moments in the athletes' journeys, offering viewers a richer understanding of the personalities involved in the sports event. The pipeline uses a combination of data retrieval from online sports databases and text-to-image generation models to create these visual biographies.

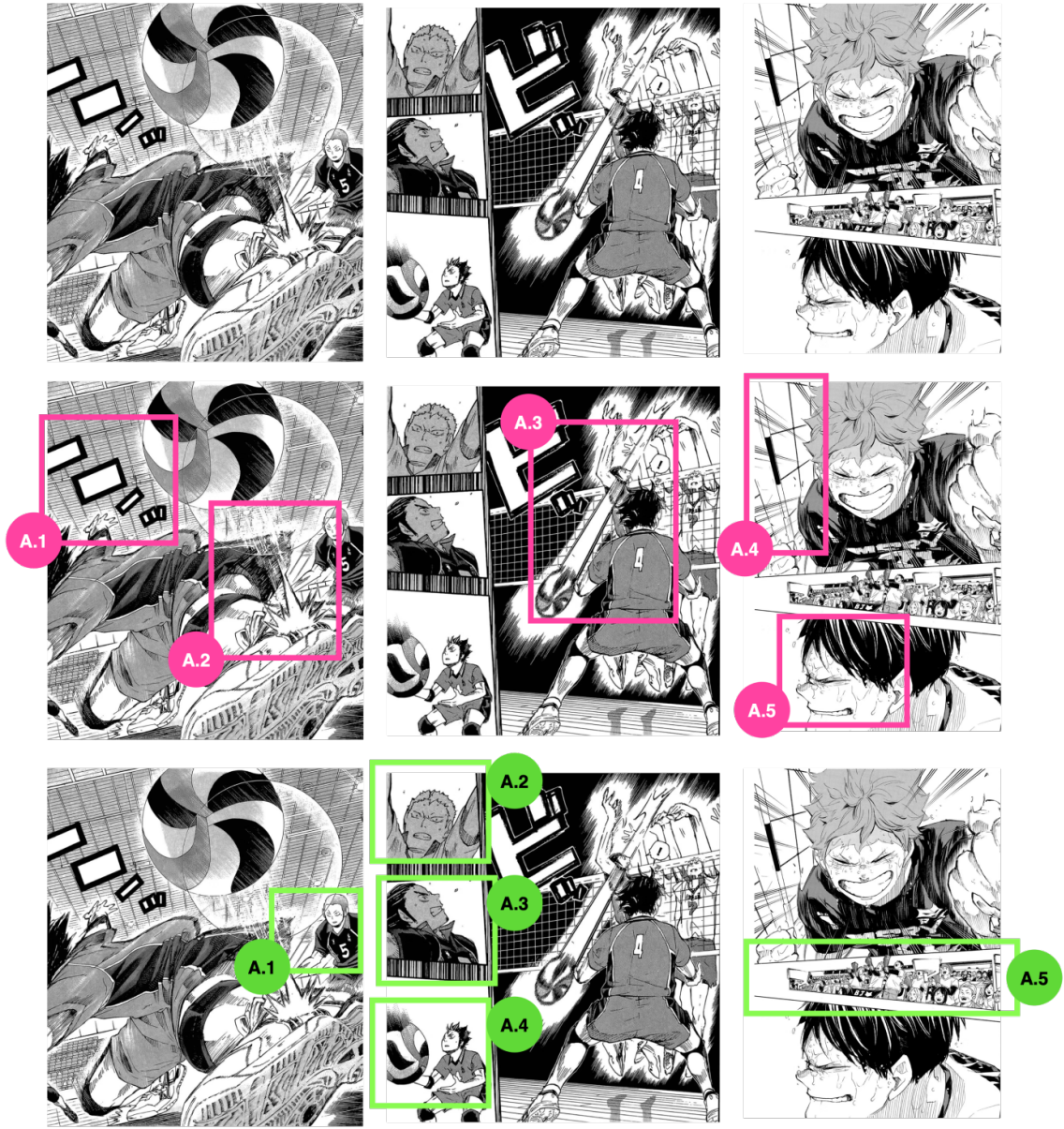

Fig. 3. Examples of Visual Elements in Sports Manga Used as Training Dataset for Style-transfer Model. The pink areas are visual hallmarks, including: (A.1) Written sound effects, (A.2) Shock wave, (A.3) Unnatural elongation of the ball to indicate motion, (A.4) Speed Lines, (A.5) Perspiration; The green areas are contextual elements, include: (A.1) Teammate1, (A.2) Competitor, (A.3) Teammate2, (A.4) Player, (A.5) Audience;

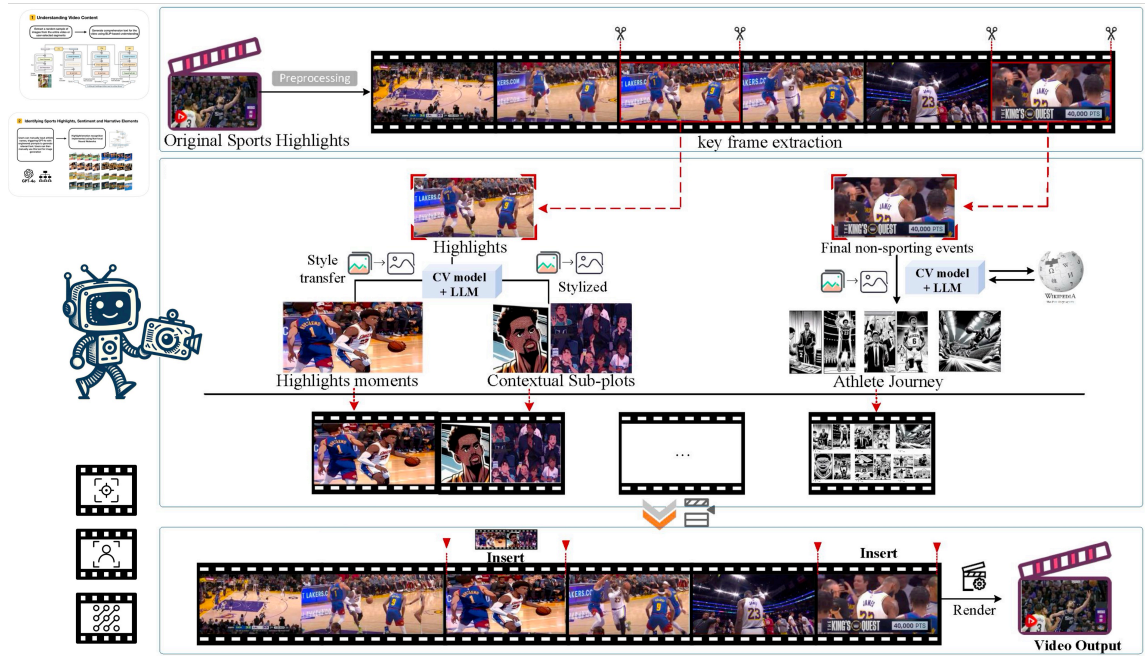

Fig. 4. Example of generation pipeline with Input Basketball Video.

*4.5.5* **[T3] Contextual Information.** The pipeline also generates additional manga-style illustrations to provide necessary contextual information. This may include visualizations of close-up shots of key stakeholders' reactions (such as teammates, coaches, spectators and live audience), to strengthen the coherence and appeal of the story. These visualizations are likely created using a combination of video understanding and style transfer algorithms to maintain the manga aesthetic.

The insertion of these generated *Manga B-roll* elements into the original video is initially automated, with the pipeline distributing the images evenly across identified gaps based on the duration of these gaps and the number of generated images. Each image is initially allocated an equal portion of the available time. However, to ensure optimal results, the pipeline provides users with the ability to manually adjust the duration of each inserted B-roll image within the system interface. This flexibility allows for fine-tuning of the pacing and emphasis of the narrative, enabling users to achieve the desired emotional impact and storytelling effect. Throughout this process, the pipeline employs appropriate transition effects, likely using techniques such as cross-fades, wipes, or more complex animated transitions, to ensure narrative fluency, creating a seamless integration of the *Manga B-roll* with the original video content and maintaining visual and emotional consistency in the enhanced video narrative.

### 4.6 Example Usage Scenarios

To obtain *Manga B-roll* sports highlight frames from the video, the user first inputs the video into the pipeline. They can then adjust the default setting by editing the provided parameters. For example, the user can adjust the video play speed or change prompts such as Manga Roll color style to generate personal preference for black and white style content. They can also input the names of sports stars appearing in the videos, generate special *Manga B-Roll* content concerning the player appearing in the video, and embed it into the consecutive sports highlight frames.

### 4.7 Generation Outcomes

To demonstrate the efficacy of our automated pipeline, we have generated 20 embedded *Manga B-Roll* videos from sports highlights from existing sports analysis datasets, such as MultiSports[46] and SportsSloMo[10] and video-sharing platforms, such as YouTube and TikTok. The sports we chose include ball sports (such as basketball, football, and volleyball), water sports (such as dividing and swimming), track and field gymnastics, and figure skating. We use videos with both horizontal and vertical aspect ratios in each type of sport. The results presented are all generated by our automated process, and there is no human authoring in this process. This is to test the effectiveness of the pipeline in different sports genres so that we can further optimize it. In the results, we focus on the generated *Manga B-Roll* and the final rendered video effects.

We selected five videos as examples of generating performance and challenges, including three in landscape orientation and two in portrait orientation. All examples can see in section 11.2 Appendix, the Figure 9, Figure 10, Figure 11, Figure 12, Figure 13. From the above results, it can be observed that our pipeline can automatically generate *Manga B-Roll* content based on the input original video and combine it to create a new sports highlight video with rich narrative and visual experience. There are two main challenges we encountered in the process of video generation, namely video quality and bias. In terms of the challenges caused by video quality, it is difficult to extract keyframes with high readability as resources for image generation due to blurred and low-quality image resolution or excessive motion speed. For example, in running videos (can see in Figure 13), it is difficult to extract key frames due to the athletes' rapid movement speed and the compression of videos on social media. As a result, generating high-quality freeze frames becomes challenging, hence leading to a smaller quantity of such frames. In addition, bias is the biggest challenge in generating consistent images of specific movements and individuals. There is a certain probability of bias in consecutive generation tasks. This may occur in all types of outcomes, particularly in career manga of athletes with a strong continuity.

### 4.8 Technical Evaluation and Iteration

We conducted technical evaluations and quality assessments on the results generated by our auto pipeline in subsection 4.7. Our tests covered a wide range of sports categories using sports highlights videos as input. The assessment primarily involved a two-part questionnaire focusing on the overall rendered video quality and specific *Manga B-roll* elements. We also conducted open-ended interviews to gather insights on system usability. Our evaluation results provided valuable insights into the system's performance. Participants generally responded positively to the visual integration and narrative enhancements offered by the *Manga B-roll* elements. However, we also identified areas for potential improvement, particularly in minimizing visual interference. These findings will guide our future refinements of the system. Detailed evaluation criteria, sports categories and results are provided in Appendix subsection 11.3. According to the results, our iterative decisions are twofold:

- Supporting more manual input options for interacting with system components. The system can become more like a toolbox for arbitrary users. For example, instead of automatically identifying key moments in the video, the arbitrary users could manually pick out moments and use the appropriate generation techniques in pipeline to inject B-rolls. This approach would give users more control over the narraitve and timing in the video and also supports for content creativity.
- Providing different level of proactive suggestions for sports highlights narration. This system offers a wider variety of automated processing options to help users select the level of AI tool assistance based on their individual needs **(DC4)**.

## 5 SPORTOONIZER: SYSTEM OVERVIEW

We implemented *Sportoonizer* as a proof-of-concept system to demonstrate and evaluate our *Manga B-roll* n-inspired approach to support augmenting sports highlights' narration and visual impact.According to the design considerations in subsection 3.2 and iterative decisions, *Sportoonizer* takes inputs sports videos to drive the auto-generation of *Manga B-roll* for narration and visual compelling **(DC1)** and **(DC2)**, user can flexibly adjust A/B roll content on the timeline to achieve control over narrative order and the construction of narrative focus **(DC3)**. The system offers users three levels of suggestions and supports manual selection of specific video clips by users to generate corresponding *Manga B-rolls* for narrative and visual creation **(DC4)**.

### 5.1 Interface

Follow the refinement of *Sportoonizer's* auto pipeline and manual manipulation supports, we developed a prototype had four primary sections (can see in Figure 5) include **1) Video Content**, which showcases the complete video content, visually depicting the interactive frame elements within the multi-track video stream; **2) Admin Control**, which is used to provide users with various degrees of AI proactive suggestions. Additionally, users can manually input and adjust the generation parameters for *Manga B-roll*; **3) Editable Timeline**, which comprises four types of video content tracks, displaying the source input video (referred to as A Roll by default) at the top, followed by the three generated *Manga B-rolls* and their respective insertion points from top to bottom. Users can view the generated content and freely adjust the arrangement order of existing video clips and video effects; **4) AI suggestions**, which displays a library of generated *Manga B-roll* content for users, facilitating their drag-and-drop into the video track for editing.

### 5.2 Implementation

The entire *Sportoonizer* system is built on CcClip framework[3]. An open-source, pure front-end audio and video editing software built with Vue3 and ffmpeg, offering all the essential features found in traditional video editing tools. We have integrated the Blip model, Shot Boundaries, and our trained highlight and sentiment recognition I3D-NL models, along with APIs from GPT-4o and Yijian[4], to provide *Sportoonizer* with a comprehensive set of capabilities. Building on the full range of features found in traditional video editors, this enables users to freely edit videos and create personalized B-roll content.

## 6 USER EVALUATION

To comprehensively evaluate the new transmedia narrative approach of *Manga B-roll* and the usability of the system with its built-in *Manga B-roll* automatic generation pipeline, our user evaluation consisted of two parts: 1) an Online Survey, and 2) a Lab study.

---

[3]CcClip: https://github.com/Cc-Edit/CcClip
[4]Yijian: https://www.yjai.art/

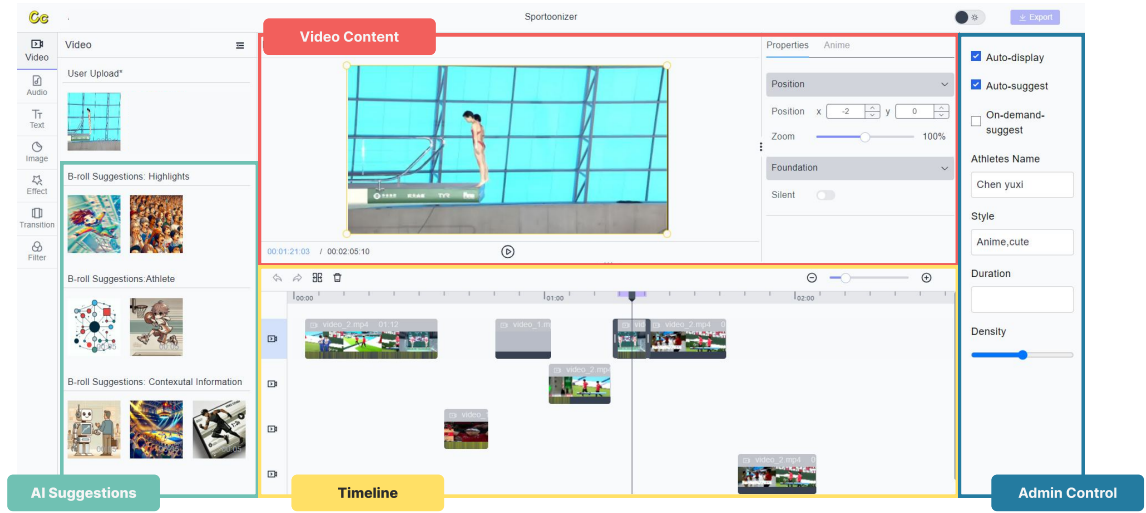

Fig. 5. *Sportoonizer* Interface which include video content in the middle area, admin control with multiple threshold adjustments in the right. And editable time in the bottom with AI suggestions about *Manga B-roll* in the left.

## 6.1 Study 1: Comparison to Ablated *Manga B-Roll* Conditions

We conducted an online survey to evaluate the quality of videos generated from our automated pipeline and their effectiveness in enhancing viewer engagement. Specifically, we aimed to assess how well the video achieved the three main objectives of the generated *Manga B-Roll*: i) developing video narration, ii) visual impact enhancement, and iii) manga-style ingenious combination of textual and non-textual elements for augmenting sports highlights under different video conditions.

*6.1.1 Materials.* To examine the research questions, for a given input sports highlights video, we generated 5 types of outcome as conditions for comparison: 1) a *Baseline* condition with the original sports highlights content, 2) a *Full* condition generated by our automated pipeline, and three ablative conditions, 3) *Narration Ablated*, 4) *Visualization Ablated*, and 5) *Integration Ablated*, generated by pipelines in which we ablated certain components based on relevant design goals.

- The *Baseline* condition uses the original videos of sports highlights, with each video containing high moments of athletes and different sports activities. The videos remain unedited, maintaining their original short form for viewing.
- The *Narration Ablated* condition does not incorporate design considerations related to narrative content formulating. In this condition, the videos do not include narrative content generated from social media data, specifically excluding B-Roll content, which is similar to the Baseline condition. We only focus on creating manga-styled highlights moments generated from sports highlights to create visual impact in the video. At the same time, we use seamless transitions and compress long shots and character action speed changes.
- The *Visualization Ablated* condition does not implement design consideration on creating manga-style visual effect. In this condition, the videos do not apply any visual effect rendered with the manga-style of the original video content. We only insert B-Roll content without any stylist transfer to complete the narrative, and it uses a realistic style of real people to supplement the flashback storyline. At the same time, we use seamless transitions and compress long shots and character action speed changes.
- The *Integration Ablated* condition does not involve design consideration on seamlessly integrating multiple media. In this condition, the videos do not achieve multimedia integration and create a sense of rhythm through seamless transitions, long shot compression, and character movements at variable speeds by any editing means. We inserted a *Manga B-roll* to meet the purposes of narratives and creating visual impact. The A-roll and B-roll content of the final video are blended together in a uniform playback.

We selected 10 input sports highlights videos from subsection 4.7 for evaluation, resulting in a total of 50 videos generated. The types of sports videos mainly include basketball, tennis, gymnastics, diving, and ice skating. For each sport type, we chose one horizontally shot video and one vertically shot video, aiming to evaluate how our pipeline performs with videos recorded in landscape and portrait orientations. The generated videos vary in length from 30 to 90 seconds.

*6.1.2 Procedure.* We conducted an online survey by recruiting participants from sports networks, social media, and video-sharing platforms. Each survey randomly selected five conditions from a pool of 10 videos in our assessment set. The order in which the five conditions were viewed was randomized to reduce the learning effect. After viewing each video, participants were asked to complete: 1) the Narrative Engagement Scale [6], which assessed narrative understanding, cognitive perspective-taking, narrative engagement, and emotional engagement, using a 7-point Likert scale ranging from 1 (strongly disagree) to 7 (strongly agree). This scale has proven effective in both video and audio narrative scenarios. 2) the Visual Impact Scale [80], which evaluates visual, emotional, and shape perception, and personal preference regarding video visual effects and different content, also using a 7-point Likert scale survey. This scale was adopted to assess visual perception responses in social media video communication. 3) The Self-defined 7-point Likert scale survey about the overall narrative and visuals of the *Sportoonizer* to understand user perceptions of the pipeline:

(1) The narration in the video is easy to understand.
(2) The visual content in the video is emotionally expressive.
(3) The integrated elements were blended smoothly into the video rhythmically.
(4) The content in the video is interesting and compelling.
(5) The viewer is provided with a rich and excellent overall viewing experience.

At the end of the survey, we also collected the demographic information of the participants and their open-ended feedback and opinions to gain a deeper understanding of users.

Afterward, we randomly selected 6 participants for post-study interviews to find out: 1) the quality and role of *Manga B-Roll* in sports highlight videos and what positive or negative effects it has compared to the original video; 2) the viewer's perception of the design consideration of narrative and visual impact, such as which narrative plots or visual elements are accessible to stimulate the audience's sense of engagement, and which ideas have a negative impact; 3) the impact on the overall viewing experience and engagement of sports highlight videos from adding *Manga B-Roll*; 5) suggestions for improving storytelling and visual impact in sports highlight videos through such transmedia elements.

### 6.2 Study 2: Understanding Users' Experience of *Sportoonizer* System

To investigate the usefulness of the system for augmenting sports highlights narration and visual compelling, we conducted a user study with 12 participants to evaluate *Sportoonizer* system. We used mixed-methods study design to gather both data from participants' survey responses and semi-structured interviews. The study examined the following two research questions:

- RQ1: How do people use the *Manga B-roll* integrated system in sports highlights narration design?
- RQ2: How do people use the *Manga B-roll* integrated system in sports highlights visual impact design?
- RQ3: What are the preferences, challenges and suggestions people face when working with *Sportoonizer* on storytelling and visual enhancements?

*6.2.1 Participants.* As our work targeted arbitrary users, we recruited twelve participants (six females and six males) from local universities, comprising six novices in content creation and six professional video content creators. The participants had diverse backgrounds and varying experience with generative AI. All were regular consumers of sports videos, watching them multiple times per week. This selection aimed to represent a broad range of potential users for the target application. Detailed demographic can see in Table 3

*6.2.2 Procedure.* After a training session to introduce *Sportoonizer*'s interface and key functionalities, participants were asked to complete a sports highlights' narration and visual impact augmenting task and output their design outcomes. The task provided video materials for the baseline condition. The sports highlight video for this task needed to include characters, plot, climax, and follow the classic Three-Act Structure[5], widely adopted in novel narratives, video production, and filmmaking. Given *Sportoonizer*'s innovative nature, no restrictions were placed on its use or completion time. We employed a think-aloud method during the experiment to gain deeper insights into the system's practicality and narrative support capabilities. With consent, we recorded each participant's entire design process for subsequent analysis. After completing the design task, we conducted a summative interview to collect feedback on the system. The interviews were semi-structured and lasted approximately 30 minutes.

### 6.3 Data Collection and Analysis

In terms of online survey in Study 1, we collected 155 valid responses to the online survey after removing duplicated answers. The participants (75 female, 78 male, 2 prefer not to say, mean age 22 years old, SD=2.15). In this study, we aggregated and standardized the data from three major categories of questions in the survey questionnaire, ranging from 1 to 7, for analysis. After data processing, which resulted in characteristics approximating a normality test, we employed the mean and standard deviation to describe the data distribution. Given that the collected data originated from repeated measurements in five different contexts, we utilized Friedman's test, a repeated measures test, to assess the potential impact of these contexts on the survey results. The data made is available at subsection 11.7. Through this approach, we could more accurately identify the specific conditions in which significant differences in audience ratings exist, providing more reliable evidence to support our research. In addition, we collected post-interview (N=6, 3 female and 3 male) audio data and transcribed them into text for open-ended analysis. In terms of user study in study 2, we collect 12 participants feedback (6 female and 6 male) about each feature of the system, narration creating and visual compelling making for sports highlights. All experimental feedback data were converted into text, with 20%

---

[5]Three-act structure: is a model used in narrative fiction that divides a story into three parts (acts), often called the Setup, the Confrontation, and the Resolution. https://en.wikipedia.org/wiki/Three-act structure

of the content independently open-coded by two authors, followed by collaborative multi-round discussions to reach consensus on themes before proceeding with the analysis of subsequent data.

## 7 FINDINGS

We demonstrate the evaluation experiment results (study 1 and study 2) through quantitative and qualitative analysis.

### 7.1 Quantitative Results

With a significance level set at $p < 0.05$, we observed significant differences in Narration Engagement and Self-defined Viewer Experience across the five conditions. To further explore the specific sources of these differences, we conducted post-hoc pairwise comparisons for Narration Engagement and Self-defined Viewer Experience, which displayed significant differences. In this process, we employed the paired Wilcoxon signed-rank test for group comparisons and applied the Bonferroni correction method to adjust for the risk of Type I errors that may arise from multiple comparisons. The results can see in Figure 14 Figure 6 and Figure 7. The study results showed that there are significant differences in Narrative Engagement and Self-defined Viewer Experience among the five conditions. The findings from the pairwise comparisons using the Bonferroni correction revealed specific differences in Narrative Engagement between the Full Condition and Baseline Condition ($P = 0.008$), Full and Visualization Ablated ($P = 0.012$), and Full and Integration Ablated ($P = 0.026$). Similarly, there was a difference in Self-defined Viewer Experience between Full Condition and Integration Ablated ($P = 0.006$).

The results suggest that there are intergroup differences in narrative presence and emotional engagement across the five conditions in the four dimensions of narrative engagement. Further pairwise comparisons using the Bonferroni correction revealed specific differences in Narrative Presence between the Full Condition and Baseline condition ($P = 0.016$), Full condition and Visualization Ablated ($P = 0.045$), and Full condition and Integration Ablated ($P = 0.012$). As for Emotional Engagement, there were differences between the Full Condition and Baseline condition ($P = 0.004$), Full condition and Visualization Ablated ($P = 0.002$), and the Full condition and Integration Ablated ($P = 0.044$).

Significant differences were observed across the five conditions in the four dimensions of Visual Impact, specifically Visual Perception. Further pairwise comparisons using the Bonferroni correction revealed specific differences in Visual Perception between the Full Condition and Visualization Ablated ($P = 0.004$), and between Full Condition and Integration Ablated ($P = 0.01$).

The five dimensions of Self-defined Viewer Experience, compelling content, and Rich Experience demonstrate intergroup differences among the five conditions. After further pairwise Bonferroni corrections, it was established that for Content Compelling, there is a difference between the Full condition and Integration Ablated ($P=0.012$). Regarding Rich View-Experience, the Full condition does not differ significantly from the other four conditions. At the same time, significant differences are observed between the Baseline condition and Visualization Ablated ($P=0.001$), Baseline condition and Integration Ablated ($P=0.028$), and Narration Ablated with Visualization Ablated ($P=0.012$). These findings can help researchers and practitioners better understand how to create more effective and engaging narratives.

### 7.2 Qualitative Results

In the online survey and the post-interview, we mainly obtain 1) users' feedback regarding their overall viewing experience of the rendered video; 2) their experience and perception of *Manga B-Roll*; and 3) their suggestions for the use of manga-style transmedia narratives in sports highlights through post-interviews. In the lab study, we main obtain

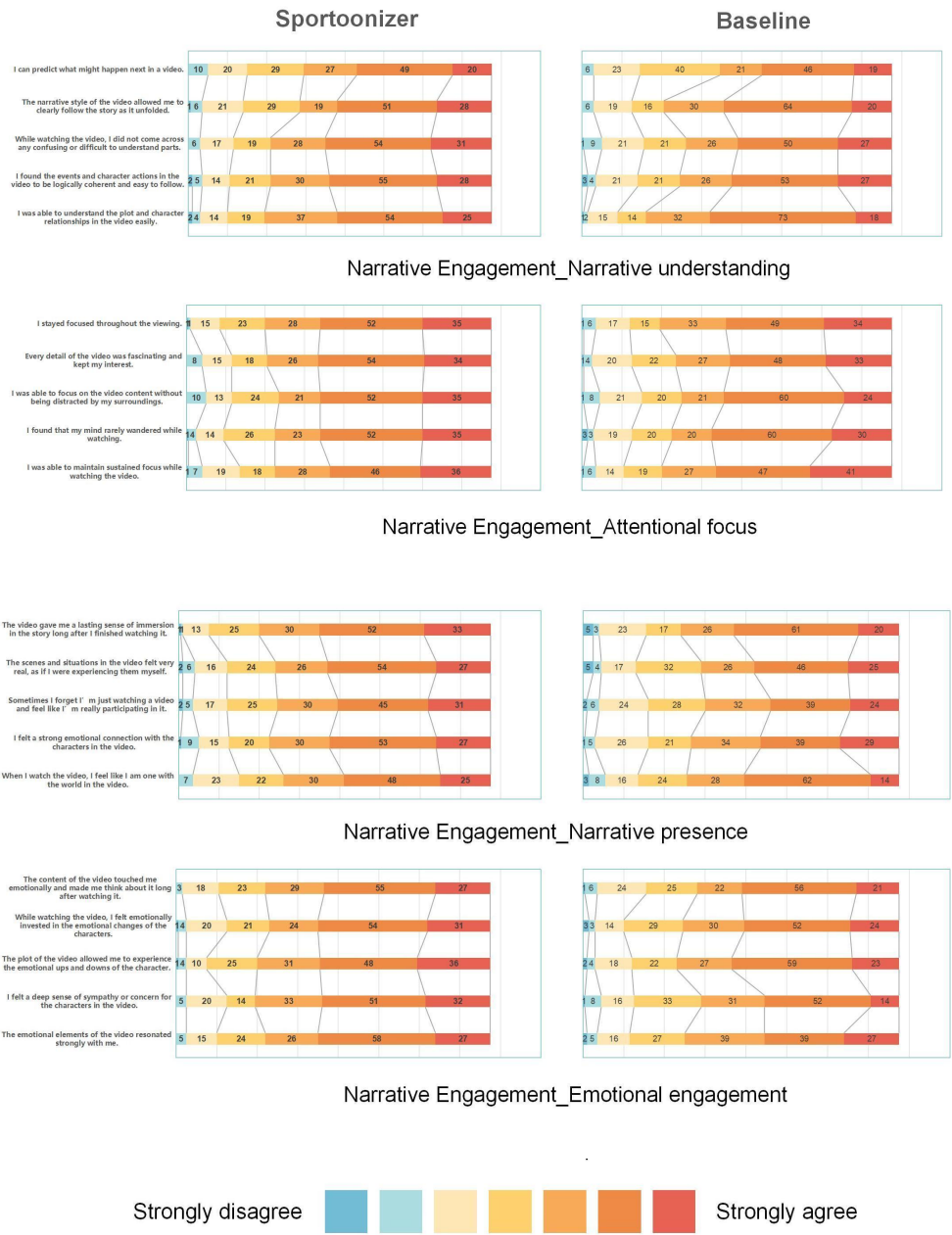

Fig. 6. Online Survey Statistics about Narrative Engagement Comparison between Full condition in *Sportoonizer* and Baseline condition

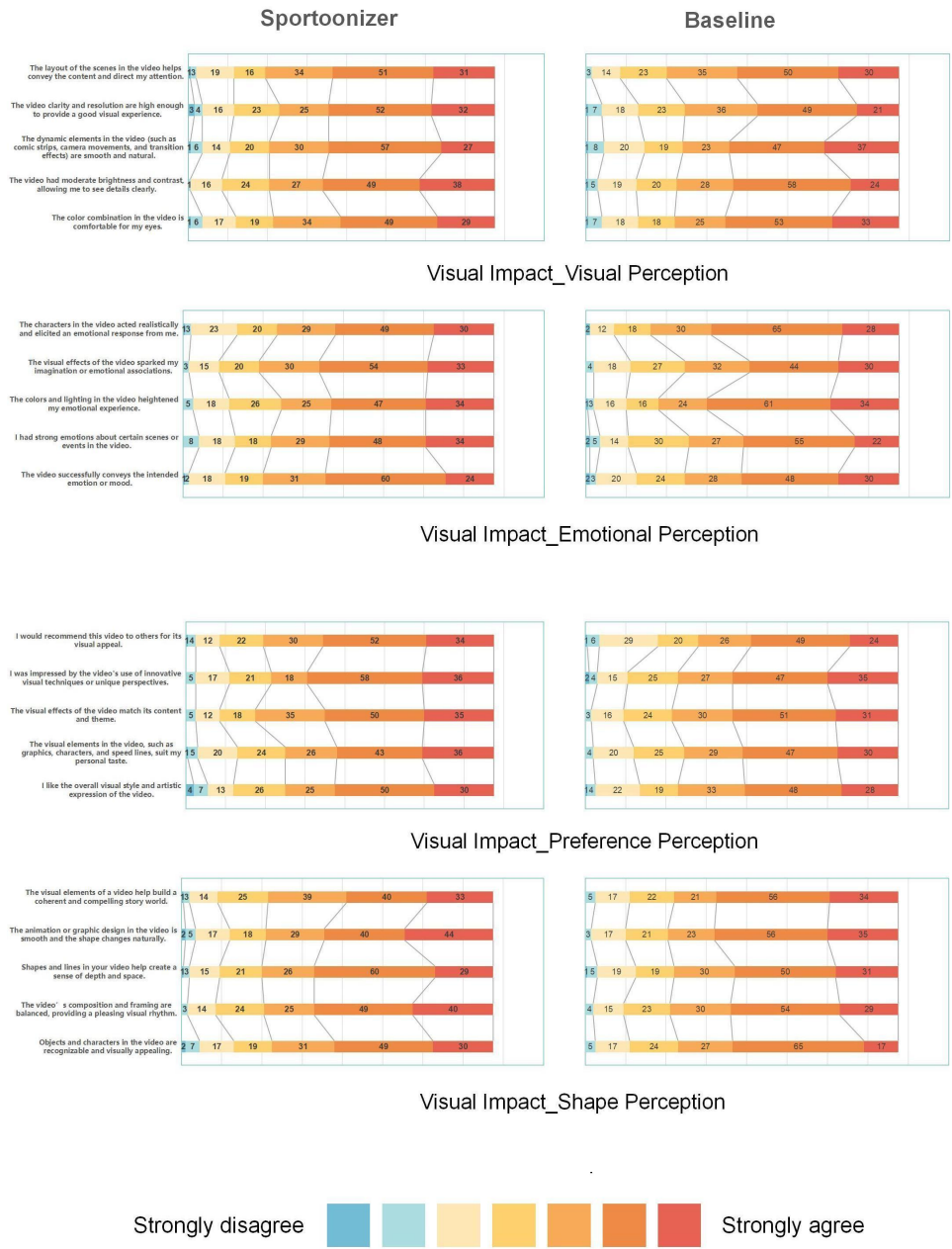

Fig. 7. Online Survey Statistics about Visual Impact Comparison between Full condition in *Sportoonizer* and Baseline condition

1) user's behaviors in employing the system for narrative and visual enhancement; and 2) perceptions of the system, encompassing preferences, challenges, and suggestions.

*7.2.1 Perception of the Rendered Video.* After viewing the video, opinions among viewers were mixed. Some praised the innovative blend of manga elements with sports footage, finding it a refreshing change that enhanced the entertainment value and potentially broadened the audience. This approach was seen as surpassing traditional sports highlights by adding a new layer of engagement. As P1 said,

> *"It's different from what I usually see on televised broadcasts, where that traditional slow play of sports is functional, and this comic style distinguishes it from traditional sports clips and traditional experiences. It's more on the entertainment side. For example, the mainstream sports video is multi-view and slow play, slow play is a continuous action, like the skating kind of multi-view slow motion, breaking the body limit is not. This gives me the feeling that just from the current scene, the key frame highlights the fixed scene of that moment of human action, fixed a moment under an angle clip, if not deliberately guide the audience, can not be noticed."*

However, we found some concerns about the manga interludes disrupting the flow of the sports action. As P4 expressed *"The quality is average, the details can't be deducted, it's easy to break the immersion and disrupt the rhythm of my watching videos."*. Critics argued that these breaks could detract from the immersive experience of traditional sports coverage by interrupting the continuity and narrative of the game, leading to a less satisfying viewing experience for those seeking authenticity.

In terms of the incorporation of manga for narration, we found this approach was mostly recognized as an innovative storytelling method that added a fresh perspective to sports events. While some viewers appreciated this novel approach, others felt it might detract from the sport itself, disrupting the natural rhythm and flow of athletic performance. As P6 said *"However, watching videos is done in order to pay attention to the game and the actions, which makes it feel more uncomfortable. It creates a disjointed style, like a sudden pop-up ad that occupies too much of my field of vision and interferes with my visual attention."* Visually, the *Manga B-roll* was noted for enhancing key sports actions and adding artistic flair to make moments more vivid and memorable. However, some critiqued the AI-generated aspects of the *Manga B-roll*, feeling it compromised the sensory experience and authenticity. P2 said, *"It is clearly visible that it is generated by AI, affecting my senses. This generated content lacks discernibility and personal style. For example, it is obvious that a cartoon style is used to generate the appearance of a real person, rather than being original cartoon content. It is very rigid."*.

All the participants thought the integration of manga with live-action footage marked a significant departure from conventional sports broadcasting, highlighting specific moments from a new perspective. While this creative narrative was unique, it also sparked concerns about losing the dynamic action and fluidity of sports. For example, P5 said *"It's like a human intervention, which will be more impressive and interrupt his continuous thinking and logic."*. Careful execution is necessary to ensure a seamless and harmonious viewing experience when combining multiple media forms.

*7.2.2 Perception of the Manga B-Roll.* The use of *Manga B-roll* in sports highlights has garnered positive and negative reactions. We found two people think it introduces a creative layer to visual storytelling, making critical moments more memorable. For example, as P1's feedback,

> *"If I just watch a continuous video, I will overlook some details or important points after watching the video, or forget or find it difficult to pay attention to the main points that the video maker wants to emphasize. But* *this B-roll gives me a sense of what the editor of the video is trying to emphasize, highlighting the highlights, and the message is something that I can absorb."*

Three participants also believed that this provided a different experience than watching the game live. For example, P3 noted, *"If it was the same as being at the venue the whole time, it would be very boring"*. However, its integration has faced criticism for noticeable style discrepancies and a lack of cohesion with live-action footage, which can disrupt the viewing experience. As P3 said, *"When I watch the game, I tend to associate it with my personal experience in sports, and B-roll interrupts personal recollections. Moreover, it is clearly an exaggerated shot that I myself cannot accomplish."*

Regarding narration, we found that some viewers thought that the *Manga B-roll* added a unique storytelling element that stands out from traditional sports broadcasting. Yet, it sometimes fragments the narrative, complicating the audience's understanding of the game's progression. As P6 said *"I am a bit unsure about where the focus should be, feeling that there are many important aspects. "*. The visual impact of the manga elements is praised for dramatizing pivotal moments, but criticisms arise over its AI-generated appearance and imperfect blending with live-action, affecting visual continuity. The continuity of aesthetic is one of the influencing factors, as indicated by P3 that *"the entertainment value of B-roll is too strong, making it easy for me to overlook or overshadow the sports aesthetic that should have been observed in the competition video."*. The novel integration of manga with live-action footage aims to enrich the viewing experience by merging different media formats.

*7.2.3 Suggestions from the Online Viewers.* Viewers offered several constructive suggestions to enhance the manga-style interludes in sports highlights. They recommended that the manga elements should be designed to seamlessly integrate with the live-action footage, with improved attention to detail and consistency in style and color. One specific suggestion was to draw inspiration from well-known sports manga like "Slam Dunk" or "Haikyuu," which could increase the recognizability and emotional connection of the *Manga B-roll*, making it more engaging and familiar to the audience.

Furthermore, viewers advised using *Manga B-roll* more sparingly and strategically to avoid overwhelming the viewer, while also maintaining the integrity of the sports narrative. By striking a balance between the entertainment value of manga elements and the authenticity of the sports content, the highlights could become a more compelling and widely appreciated form of sports entertainment.

In addition to these suggestions, viewers also highlighted the importance of considering the context in which the sports highlights are presented. *Manga B-rolls* might be more acceptable for marketing purposes or casual viewing due to their entertainment nature. However, for dedicated sports enthusiasts who prioritize the integrity of the game, a more subdued approach might be preferred to avoid disrupting their viewing experience.

*7.2.4 Perception of the System.* We observed participants exhibiting diverse practices, such as selecting, generating, arranging, and embedding *Manga B-roll* into input videos to narrate and visually enhance sports highlights. The system not only aids but significantly elevates users' narrative capabilities, potentially transforming the landscape of sports highlight storytelling.

*Augmenting Narrative Design: Creative, Expressiveness and Storyline Reconstruction.* Our findings demonstrate the system's effectiveness in enhancing narrative design. Participants leveraged various features, particularly the timeline's highlight moments and generated *Manga B-roll*, to spark creative storytelling. For example, P7 utilized automatically identified key actions to craft a character-focused training narrative. As he expressed, *"His action of scoring the goal* was particularly amazing, requiring a lot of training and skill. It was this clip that made me think of creating a flashback story to express how important this moment was.".

Additionally, The generated *Manga B-roll* content proves to be a powerful catalyst for narrative innovation. This novel form of visual content stimulates users' creative thinking, encouraging them to explore diverse storytelling approaches. P4's strategy of utilizing the B-roll as an introductory background to build narrative tension exemplifies how this feature enhances storytelling depth. By enabling users to weave compelling backstories and create narrative arcs, the *Manga B-roll* significantly elevates the quality and engagement of sports highlight narration, potentially revolutionizing the field of visual storytelling. As she mentioned, *"I know athletes have it particularly tough, so I want to use this content as an introductory background, inserting it when introducing her competition entrance, as a lead-up to showing her highest score later."*

Some participants also mentioned that the system provided them with support in organizing the narrative order. Its editable interface empowers users to craft both linear and nonlinear narratives by flexibly positioning supplementary B-roll content. For example, P2 said *"Sports highlights can be used as the beginning of the story, or as the climax in the middle, or the end, and my main idea is when to insert the Manga B-roll to explain why, or to use the generated content to cover the game."*. His feedback underscores the system's versatility in storytelling, allowing strategic placement of *Manga B-roll* to either explain context or cover game action. This flexibility, coupled with the ability to select A-roll clips and generate supportive B-roll, enables users to precisely highlight story focus and enhance meaning expression. Furthermore, the system's provision of extracted key segments aids in reconstructing story arcs and enriching narrative development. Collectively, these features not only streamline the storytelling process but also substantially improve content coherence, potentially revolutionizing sports highlight narration and elevating overall viewer engagement. As P11 said:

> *"I can make this clip as a 'life flashing before one's eyes'[6] for the athlete before the end of the action. It's like the manga 'Touch'[7] who keeps reminiscing about his bat swing, then recalling his motivation for participating in the sport, and finally hitting the ball as the end of the story.".*

*Enhancing Visual Impact: Freshness, Composition, and Dynamic Elements.* Our study demonstrates that participants effectively enhanced visual appeal through the system in three key ways: by creating visual novelty, implementing immersive camera perspectives, and generating dynamic visual experiences. These improvements significantly boosted the overall quality of the sports highlight videos, showcasing the system's potential to transform visual storytelling in sports content. As one of our participant noted, *"Blending realistic styles with manga scene transitions complements each other and adds scenes impossible to film in real life. (...) Furthermore, the generated anime content includes various cinematographic techniques which, when combined with the original video, increases viewer engagement through diverse shot compositions. (P9)"*

*7.2.5 Wishlist from the System Users.* Our study reveals critical areas for enhancement in our video authoring system. While providing fundamental capabilities, the system falls short in meeting advanced user needs. Users strongly advocate for more precise B-roll placement recommendations, as the current automatic insertion often causes content overlap with the original video, necessitating time-consuming manual adjustments. Additionally, there's a notable discrepancy between user input and generated content, particularly in character and text alignment with the input video, which hampers seamless integration. Users also expressed a need for higher quality in generated *Manga B-Roll* content. They also suggest employing visual methods in the interface to: 1) distinguish between A/B-roll, 2) differentiate B-roll with different functions, and 3) support non-fine-grained timeline adjustments, collectively supporting the realization of narrative creativity.

---

[6]life flashing before one's eyes (Soumatou in comic): refers to the phenomenon where a person sees a rapid succession of memories from their life, often when facing death or during intense moments.
[7]Touch: is a Japanese high school baseball manga series written and illustrated by Mitsuru Adachi. https://en.wikipedia.org/wiki/Touch_(manga)

## 8 DISCUSSION

Our study showed that both strengths and areas for improvement in our video authoring system. The system successfully provides basic video editing capabilities and introduces innovative features like *Manga B-roll* generation, enhancing visual novelty in sports highlights. Users appreciated the creative freedom and flexibility offered by the system, allowing for unique narratives and visual styles. However, several key areas require refinement. The automatic insertion of generated content often results in overlap with the original video, necessitating manual adjustments. Users emphasized the need for more precise placement recommendations. Additionally, there's a notable discrepancy between user input and generated content, particularly in character and text alignment. The quality of generated *Manga B-roll* also needs improvement. Here, we discuss about the significance, generalizability and potentiality as matter of our research.

**Significance – augmenting narration and visual impact beyond videos.** Our system combines dynamic manga sequences with live-action footage to create emotionally immersive sports highlight videos. This innovative approach blends realism with stylized art, capturing viewers' attention and surpassing previous AI-powered content creation efforts within this domain [48, 52]. By using the visual language of mangas to reinterpret real-world sports footage, we offer a nuanced understanding of athletes' movements, emotions, background and contextual information that traditional sports commentary often misses. From the findings, most viewer stated that this style of B-roll could effectively enhance storytelling by providing additional emotional and cultural information in short sports videos,. It uses exaggerated expressions, dynamic movements, and vivid color contrasts commonly seen in cartoons to create a visually striking impact. Combined with the results of survey data analysis, our narrative ability has obvious effectiveness. This proves this style of B-roll effectively conveys the athletes' narrative emotions, the tension of the competition, and the deeper meanings of sportsmanship.

**Generalizability – extending from tested sports to other sports.** Our system explores adaptability in different sports, emphasizing flexibility and potential to meet diverse demands for sports content. Furthermore, its automated generation capability can customize manga sequences and narratives based on the characteristics and rules of different sports, enhancing the viewing experience for each sport, as the same as mentioned by previous work [11, 13]. According to user feedback, the innovative content of the system may attract audiences who have not previously been engaged, thereby expanding the appeal of sports media.

**Potentiality – leveraging LLMs to enhance narration and visual impact.** The integration of LLMs represents a valuable opportunity to enhance the system's capacity for generating dynamic and personalized sports stories using AI's powerful language generation capabilities. By seamlessly incorporating its output with other media formats, such as podcasts and social platforms, the system has the potential to create multi-dimensional sports experiences that engage audiences in constructive and meaningful ways.

## 9 LIMITATIONS AND FUTURE WORK

Our forthcoming research endeavors are geared toward refining the sports video experience. We will explore the behavioral patterns of the audience, incorporate generative AI techniques, and study the potential of comic-style aesthetics.

Our primary goal is to enhance our attention segmentation algorithms, streamline the creation and arrangement of video content, and elevate attention signals with generative AI models. Through the exploration of multifaceted multimodal approaches, we aspire to provide viewers with a more gratifying sports video experience.

## 10 CONCLUSION

In this paper, we present a new approach that merges B-roll footage with Manga style to enhance the storytelling ability and emotional conveyance of the narrative in sports videos. Our innovative automated pipeline combines *Manga B-roll* generation with original videos. With the help of *Sportoonizer*, content creators can quickly create captivating storylines into sports highlights. We evaluate the system with viewers using both qualitative and quantitative methods and demonstrate that it enhances viewer engagement with visual narratives and provides a unique and creative perspective. However, it also raises concerns about the potential loss of dynamic action and fluidity inherent in sports. Overall, our approach has the potential to revolutionize the way sports highlights are created, providing a more immersive and engaging experience for viewers.

Conference acronym 'XX, June 03–05, 2018, Woodstock, NY    Hu, et al.

## 11 APPENDIX

### 11.1 Demographic: Participatory Design Workshop

| ID | Age | Gender | Creating Experience | Education Level | Occupation | Stakeholder Type |
|---|---|---|---|---|---|---|
| P1 | 21 | Female | Somewhat | Bachelor | Pharmacist | Sports enthusiast |
| P2 | 26 | Male | Not at All | Doctoral | Student | Sports spectator |
| P3 | 19 | Male | Moderately | Bachelor | Student | Sports athlete |
| P4 | 35 | Male | Not at All | Master | Coach | Sports coach |
| P5 | 19 | Female | Somewhat | Bachelor | Student | Sports student |
| P6 | 41 | Male | Somewhat | Doctoral | Professor | Sports teacher |
| P7 | 23 | Female | Moderately | Master | Commentator | Sports commentator |
| P8 | 25 | Male | Moderately | Master | Live streamers | Sports commentator |
| P9 | 28 | Female | Extremely | Master | Influencer Blogger | Video content creator |
| P10 | 32 | Female | Extremely | Doctoral | Vlogger | Video content creator |
| P11 | 26 | Male | Extremely | Master | Video scriptwriter | Screenwriter |
| P12 | 44 | Female | Extremely | Bachelor | Video maker | Video editor |
| P13 | 23 | Male | Extremely | Bachelor | Special effect | Video technical staff |
| P14 | 17 | Male | Somewhat | High school graduate | Student | Anime enthusiast |
| P15 | 22 | Female | Somewhat | Bachelor | Student | Anime enthusiast |

Table 2. Participants Demographics include age, gender, content creating experience, education level, occupation and stakeholder type.

### 11.2 Demographic: Study 2 in User Evaluation

| ID | Age | Gender | GenAI Using Experience | Video Creating Experience |
|---|---|---|---|---|
| P1 | 33 | Male | Somewhat | Professional |
| P2 | 22 | Male | Somewhat | Novice |
| P3 | 45 | Female | Not at All | Professional |
| P4 | 38 | Female | Somewhat | Professional |
| P5 | 19 | Male | Somewhat | Novice |
| P6 | 26 | Male | Extremely | Novice |
| P7 | 42 | Female | Somewhat | Professional |
| P8 | 34 | Male | Not at All | Professional |
| P9 | 35 | Female | Moderately | Professional |
| P10 | 29 | Female | Extremely | Novice |
| P11 | 21 | Female | Moderately | Novice |
| P12 | 27 | Male | Moderately | Novice |

Table 3. Participants Demographics include age, gender, GenAI using experience, and video creating experience.

### 11.3 Detailed Technical Evaluation

Our evaluation covered the following sports categories:

- Ball sports (e.g., basketball, football, volleyball)
- Water sports (e.g., diving, swimming)
- Track and field
- Gymnastics
- Figure skating

The questionnaire consisted of two main parts:

(1) For the overall rendered video:
   - Visual coherence (style integration of Manga B-roll with original video)
   - Narrative enhancement (contribution of Manga B-roll to storytelling)
   - Information transmission efficiency (viewer comprehension and retention of additional information)
   - Viewing experience (overall comfort and appeal)
(2) Specific to Manga B-roll elements:
   - Style consistency (visual harmony among different Manga B-roll elements)
   - Timeliness of information presentation (appropriateness of Manga B-roll timing)
   - Content relevance (correlation between Manga B-roll and video content)
   - Visual interference (assessment of whether Manga B-roll affects original video viewing)

Additionally, we conducted open-ended interviews regarding the usability of Manga B-roll and the system to facilitate further iteration and optimization.

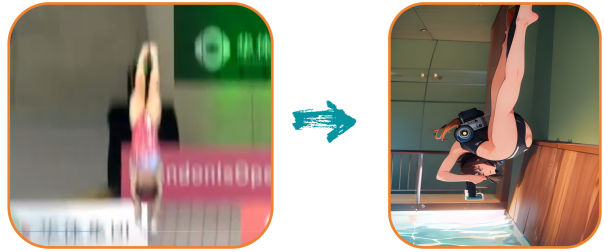

Fig. 8. Failure Generation

### 11.4 Prompt Template

*11.4.1 Prompt to Generate Highlights and Contextual Sub-plots.* I need an image in the style of Japanese manga. The relevance of the generated image should be 50% in relation to the source-image I provide. Please create one combining the image I give you and my prompt.

*11.4.2 Prompt to Generate Athlete Journey.* I want you to help me search for all the information about an {input1: the type of the sport} athlete named {input2: athlete name} from specified sources like WIKI or the official Olympic athletes' database. Then summarize his career into {input3: number of stages} stages. Just output the description texts of the {input2: athlete name} stages using English. Do not output any other content.

Do not generate any captions in all the images or pictures. The main character is an {input1: the type of the sport} athlete. Generate black-and-white manga-style comics. Use no more than two storyboards to split the content, and the two photo frames must be divided using comic dividers.

### 11.5 Failure Case

During the development and testing of our pipeline, we had difficulty in accurately capturing highlight frames when processing extremely unclear videos. This is due to the current technical limitations of computer vision (CV) models, which are unable to recognize content in highly blurred images with great precision. Consequently, when the input original video is blurry, the quality of the generated video tends to be low and lacks strong visual appeal, like Figure 10. Therefore, in the subsequent tuning process of the pipeline, we selected videos with a clarity of 1080p or higher as the test dataset.

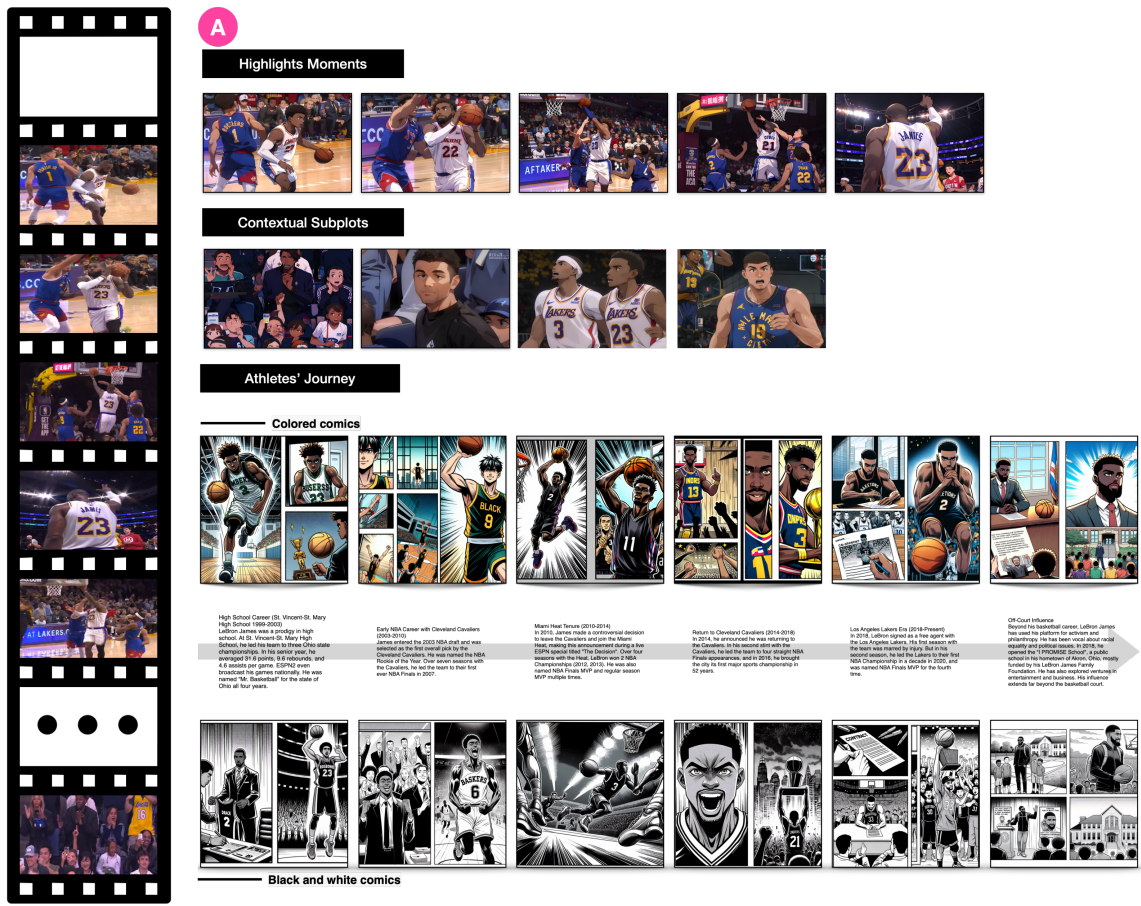

Fig. 9. We present the **_Results [A]_** automatically generated by our pipeline from inputs that are challenging to generate Manga B-Roll. Left: key-frames from the original video, Right: the content included in the generated Manga B-Roll. This is a horizontal basketball video.

## 11.6 Examples of Generation Challenges

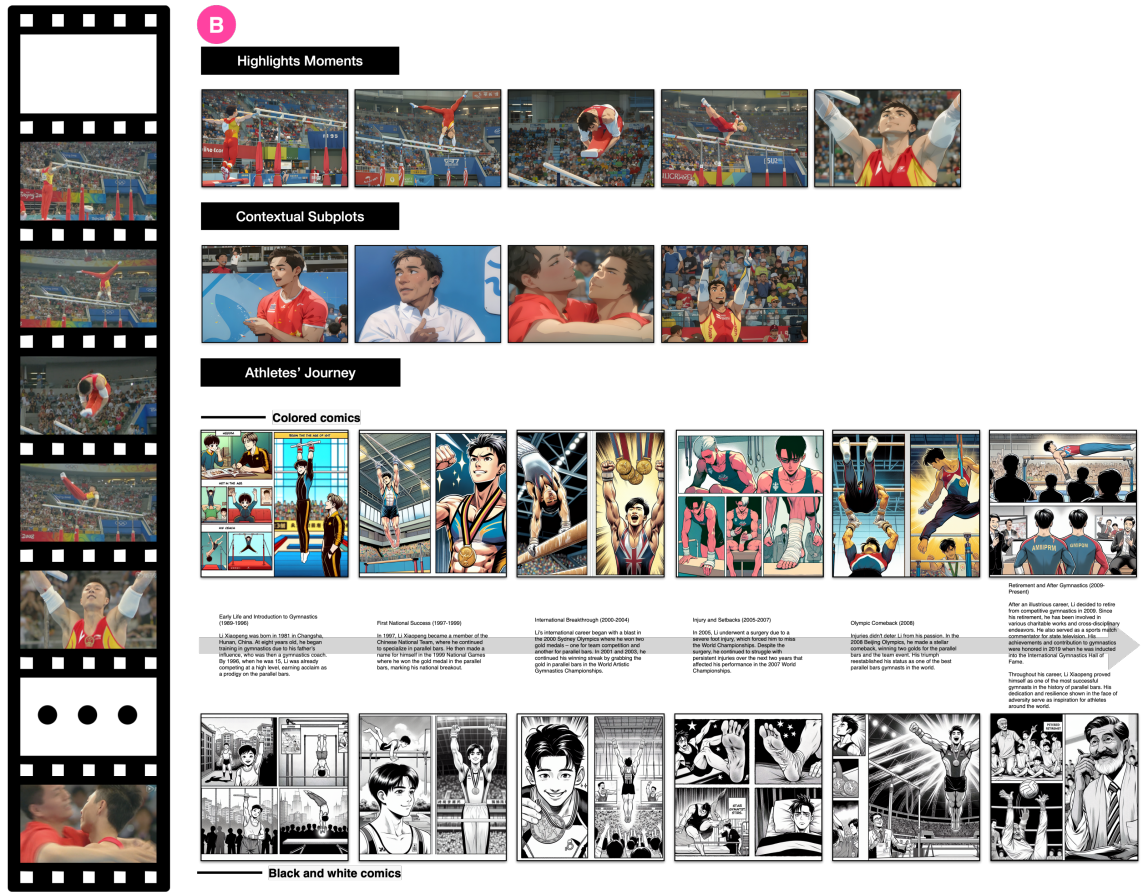

Fig. 10. We present the ***Results [B]*** automatically generated by our pipeline from inputs that are challenging to generate Manga B-Roll. Left: key-frames from the original video; Right: the content included in the generated Manga B-Roll. This is a horizontal gymnastics video.

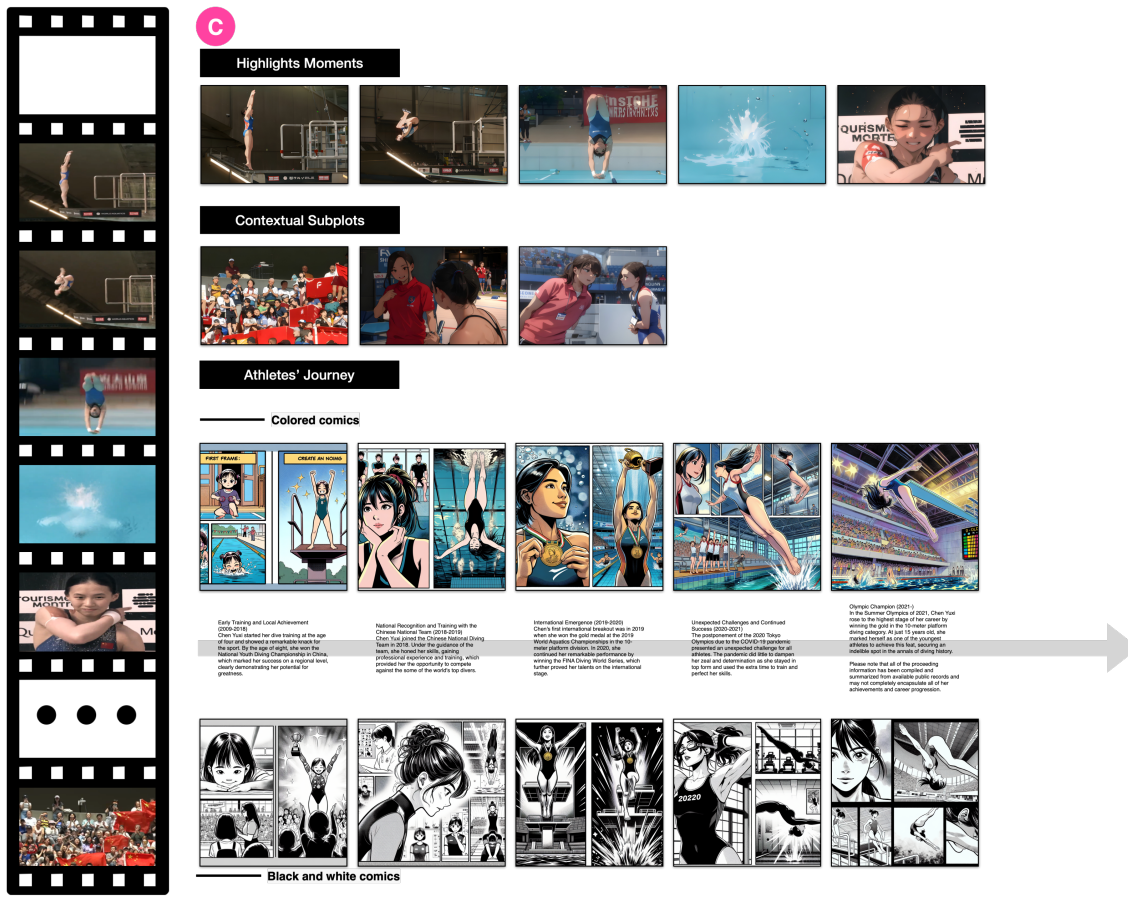

Fig. 11. We present the **Results [C]** automatically generated by our pipeline from inputs that are challenging to generate Manga B-Roll. Left: key-frames from the original video; Right: the content included in the generated Manga B-Roll. This is a horizontal diving video.

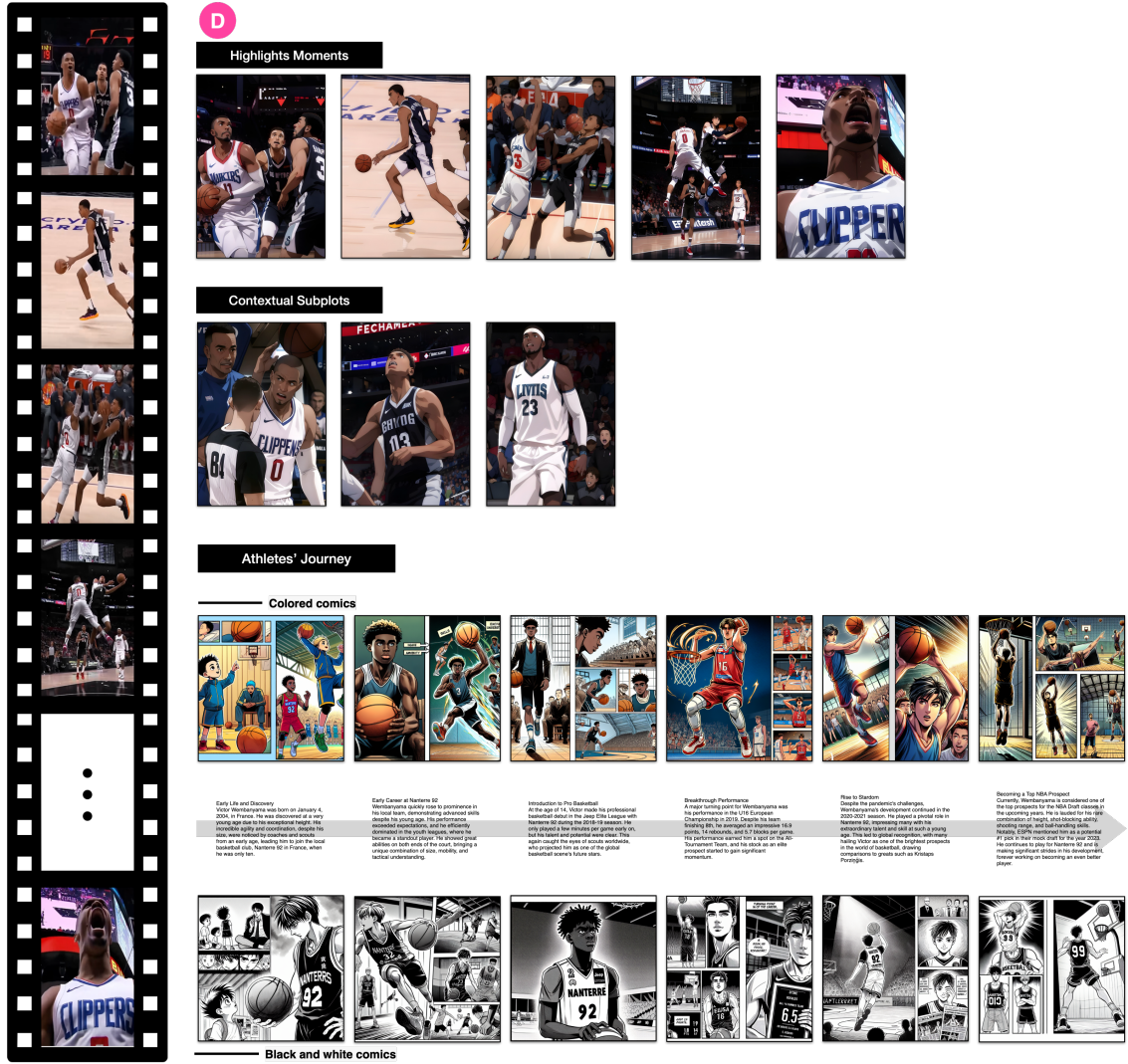

Fig. 12. We present the **Results [D]** automatically generated by our pipeline from inputs that are challenging to generate Manga B-Roll. Left: key-frames from the original video; Right: the content included in the generated Manga B-Roll. This is a vertical basketball video.

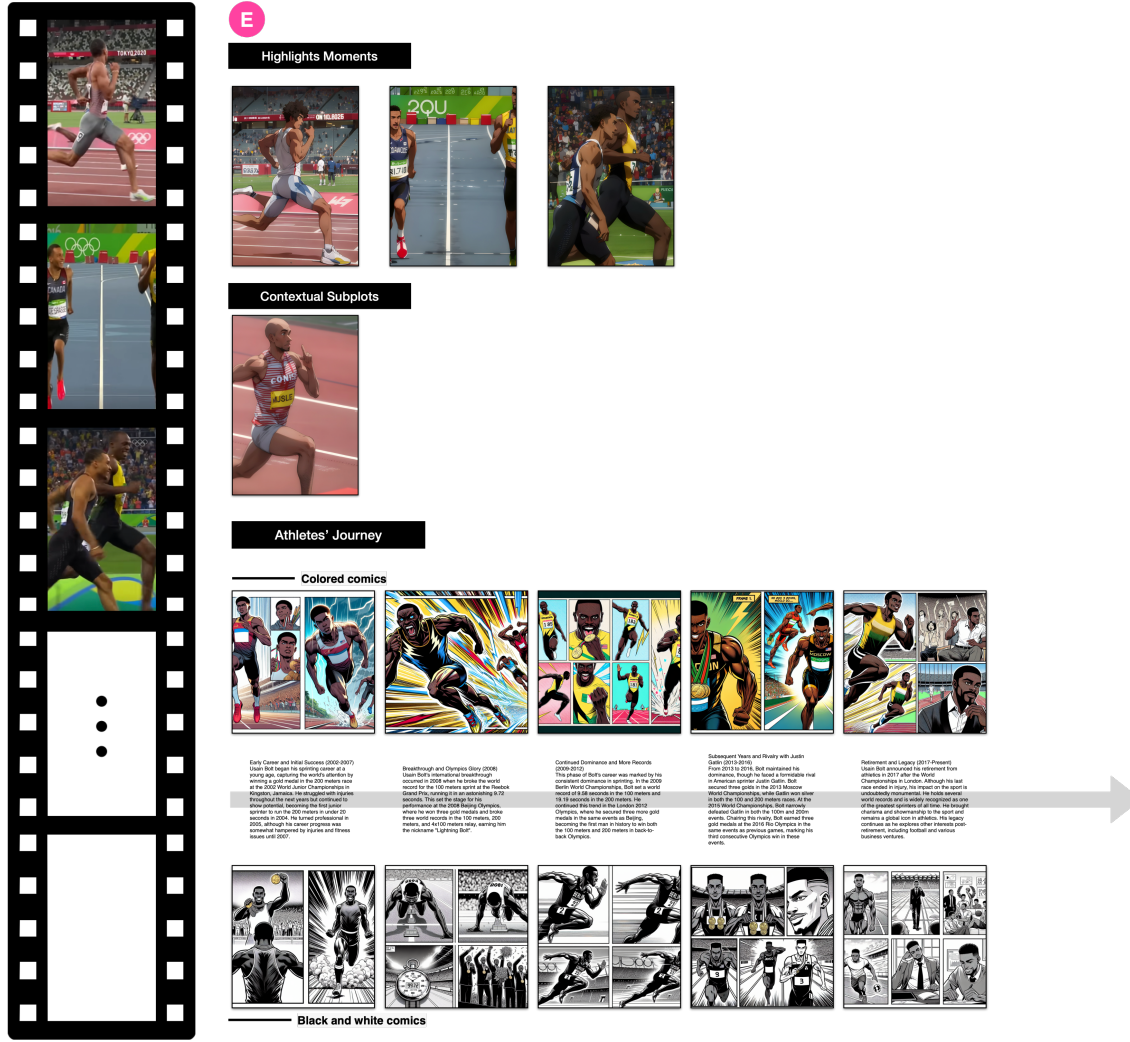

Fig. 13. We present the **Results [E]** automatically generated by our pipeline from inputs that are challenging to generate Manga B-Roll. Left: key-frames from the original video; Right: the content included in the generated Manga B-Roll. This is a vertical running video.

## 11.7 User Evaluation: Online Survey Statistics

Table 4. Statistics of the Friedman test results for inter-group differences

| Variable | Base condition | Full condition | Narration ablated | Visualization ablated | Integration ablated | Statistics | P |
| --- | --- | --- | --- | --- | --- | --- | --- |
| Narrative Engagement | 5.11 (1) | 5.24 (0.95) | 5.18 (1.01) | 5.02 (1.07) | 4.99 (1.02) | 11.981 | 0.017 |
| Visual Impact | 5.24 (0.96) | 5.29 (0.97) | 5.21 (1.02) | 5.04 (1.1) | 5.05 (1.02) | 5.238 | 0.264 |
| Viewer Experience | 5.34 (1) | 5.31 (1.07) | 5.23 (1.13) | 5.05 (1.26) | 5.04 (1.16) | 11.624 | 0.02 |

Table 5. Statistics of the Pairwise comparison Bonferroni test

| variable | Baseline Condition_ Full Condition | Full Condition_ Narration Ablated | Full Condition_ Visualization Ablated | Full Condition_ Integration Ablated | Baseline Condition_ Narration Ablated | Baseline Condition_ Visualization Ablated | Baseline Condition_ Integration Ablated | Narration Ablated_ Visualization Ablated | Narration Ablated_ Integration Ablated | Visualization Ablated_ Integration Ablated |
| --- | --- | --- | --- | --- | --- | --- | --- | --- | --- | --- |
| Narrative Engagement | 0.008 | 0.667 | 0.026 | 0.012 | 0.026 | 0.726 | 0.546 | 0.053 | 0.015 | 0.959 |
| Visual Impact | 0.416 | 0.23 | 0.032 | 0.016 | 0.993 | 0.135 | 0.134 | 0.078 | 0.017 | 0.609 |
| Viewer Experience | 0.787 | 0.857 | 0.084 | 0.006 | 0.667 | 0.006 | 0.001 | 0.058 | 0.022 | 0.58 |

Table 6. Statistics of the Friedman test results for inter-group differences

| Variable | Base condition | Full condition | Narration ablated | Visualization ablated | Integration ablated | Statistics | P |
| --- | --- | --- | --- | --- | --- | --- | --- |
| Narrative understanding | 5.13 (1) | 5.16 (1.06) | 5.09 (1.09) | 5 (1.13) | 4.96 (1.15) | 2.709 | 0.608 |
| Attentional focus | 5.26 (1.17) | 5.31 (1.1) | 5.29 (1.1) | 5.1 (1.19) | 5.09 (1.14) | 6.952 | 0.138 |
| Narrative presence | 4.99 (1.16) | 5.19 (1.08) | 5.14 (1.16) | 5 (1.19) | 4.91 (1.17) | 14.23 | 0.007 |
| Emotional engagement | 5.06 (1.11) | 5.29 (1.04) | 5.19 (1.07) | 5 (1.17) | 5.02 (1.11) | 14.155 | 0.007 |

Table 7. Statistics of the Pairwise comparison Bonferroni test

| variable | Baseline Condition_ Full Condition | Full Condition_ Narration Ablated | Full Condition_ Visualization Ablated | Full Condition_ Integration Ablated | Baseline Condition_ Narration Ablated | Baseline Condition_ Visualization Ablated | Baseline Condition_ Integration Ablated | Narration Ablated_ Visualization Ablated | Narration Ablated_ Integration Ablated | Visualization Ablated_ Integration Ablated |
| --- | --- | --- | --- | --- | --- | --- | --- | --- | --- | --- |
| Narrative understanding | 0.424 | 0.75 | 0.165 | 0.059 | 0.744 | 0.431 | 0.255 | 0.466 | 0.086 | 0.826 |
| Attentional focus | 0.435 | 0.934 | 0.081 | 0.028 | 0.362 | 0.113 | 0.119 | 0.079 | 0.02 | 0.983 |
| Narrative presence | 0.016 | 0.494 | 0.045 | 0.012 | 0.04 | 0.434 | 0.423 | 0.139 | 0.007 | 0.242 |
| Emotional engagement | 0.004 | 0.281 | 0.002 | 0.044 | 0.093 | 0.85 | 0.643 | 0.091 | 0.082 | 0.873 |

## 11.8 Online Survey: Aggregated Likert Scale Ratings

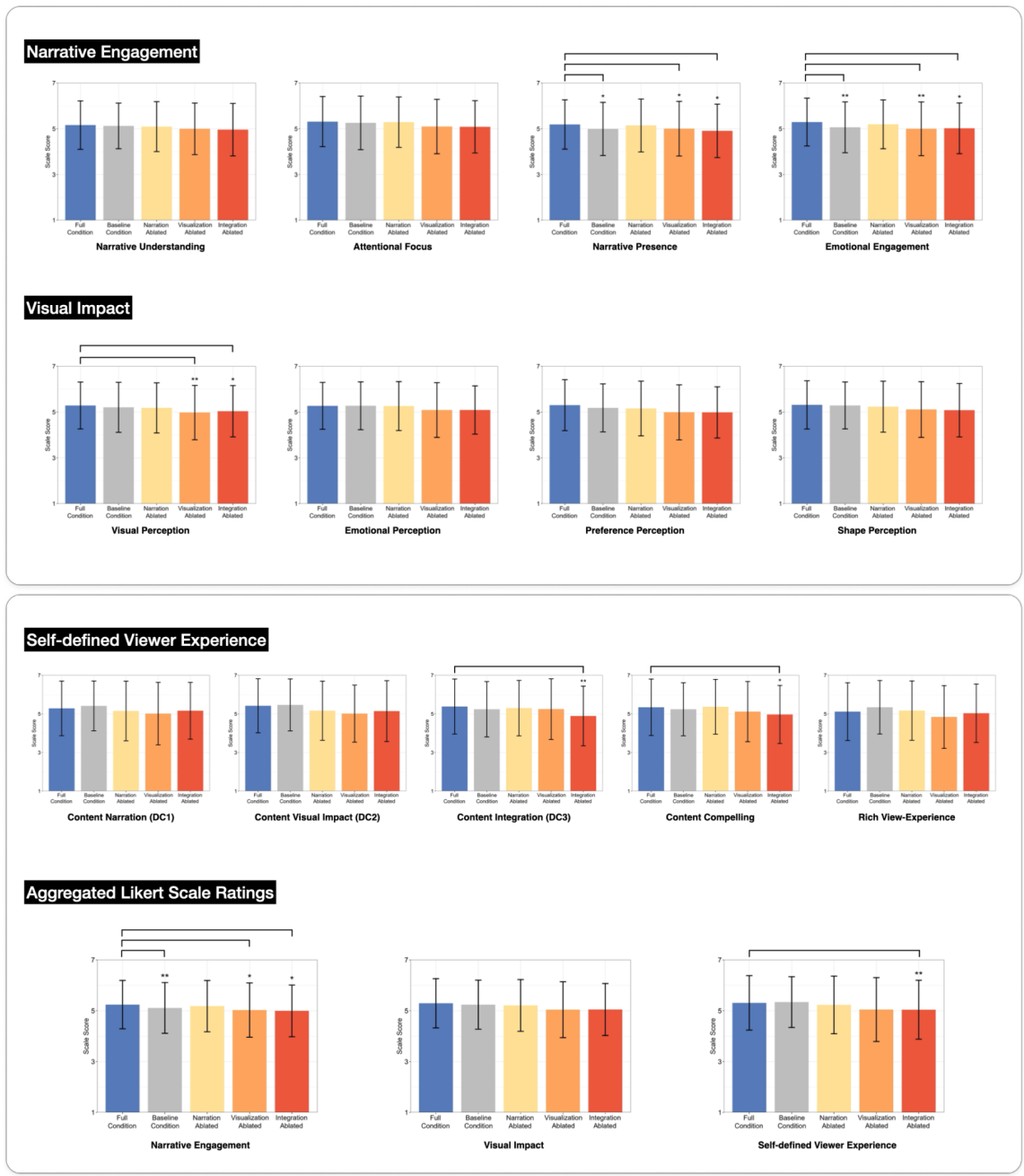

*p<.05     **p<.01     ***p<.001     ****p<.0001

Fig. 14. User Evaluate Statistics include Narrative Engagement, Visual Impact, Self-defined Viewer Experience and Aggregated Likert scale ratings of questions on Narration, Visual Impact, and Integration.

Table 8. Statistics of the Friedman test results for inter-group differences

| Variable | Base condition | Full condition | Narration ablated | Visualization ablated | Integration ablated | Statistics | P |
| --- | --- | --- | --- | --- | --- | --- | --- |
| Visual Perception | 5.21 (1.09) | 5.29 (1.03) | 5.18 (1.09) | 4.98 (1.19) | 5.04 (1.12) | 14.092 | 0.007 |
| Emotional Perception | 5.27 (1.05) | 5.27 (1.02) | 5.26 (1.07) | 5.09 (1.2) | 5.09 (1.05) | 7.071 | 0.132 |
| Preference Perception | 5.18 (1.05) | 5.3 (1.11) | 5.15 (1.19) | 4.99 (1.2) | 4.98 (1.12) | 8.872 | 0.064 |
| Shape Perception | 5.29 (1.03) | 5.31 (1.05) | 5.23 (1.11) | 5.11 (1.22) | 5.08 (1.17) | 3.17 | 0.53 |

Table 9. Statistics of the Pairwise comparison Bonferroni test

| variable | Baseline Condition_ Full Condition | Full Condition_ Narration Ablated | Full Condition_ Visualization Ablated | Full Condition_ Integration Ablated | Baseline Condition_ Narration Ablated | Baseline Condition_ Visualization Ablated | Baseline Condition_ Integration Ablated | Narration Ablated_ Visualization Ablated | Narration Ablated_ Integration Ablated | Visualization Ablated_ Integration Ablated |
| --- | --- | --- | --- | --- | --- | --- | --- | --- | --- | --- |
| Visual Perception | 0.332 | 0.291 | 0.004 | 0.01 | 0.986 | 0.022 | 0.13 | 0.016 | 0.065 | 0.345 |
| Emotional Perception | 0.886 | 0.973 | 0.057 | 0.056 | 0.839 | 0.121 | 0.041 | 0.1 | 0.052 | 0.692 |
| Preference Perception | 0.153 | 0.271 | 0.018 | 0.007 | 0.712 | 0.179 | 0.09 | 0.037 | 0.022 | 0.636 |
| Shape Perception | 0.868 | 0.606 | 0.521 | 0.11 | 0.832 | 0.222 | 0.191 | 0.317 | 0.043 | 0.513 |

Table 10. Statistics of the Friedman test results for inter-group differences

| Variable | Base condition | Full condition | Narration ablated | Visualization ablated | Integration ablated | Statistics | P |
| --- | --- | --- | --- | --- | --- | --- | --- |
| DC1 | 5.41 (1.29) | 5.28 (1.41) | 5.15 (1.54) | 5.01 (1.62) | 5.16 (1.47) | 5.535 | 0.237 |
| DC2 | 5.46 (1.34) | 5.42 (1.4) | 5.16 (1.53) | 5.01 (1.48) | 5.14 (1.58) | 6.713 | 0.152 |
| DC3 | 5.24 (1.43) | 5.38 (1.42) | 5.3 (1.43) | 5.25 (1.57) | 4.89 (1.54) | 8.575 | 0.073 |
| Compelling | 5.24 (1.37) | 5.34 (1.46) | 5.37 (1.42) | 5.12 (1.56) | 4.97 (1.51) | 10.305 | 0.036 |
| Rich Experience | 5.34 (1.38) | 5.12 (1.49) | 5.17 (1.54) | 4.84 (1.63) | 5.03 (1.51) | 10.081 | 0.039 |

Table 11. Statistics of the Pairwise comparison Bonferroni test

| variable | Baseline Condition_ Full Condition | Full Condition_ Narration Ablated | Full Condition_ Visualization Ablated | Full Condition_ Integration Ablated | Baseline Condition_ Narration Ablated | Baseline Condition_ Visualization Ablated | Baseline Condition_ Integration Ablated | Narration Ablated_ Visualization Ablated | Narration Ablated_ Integration Ablated | Visualization Ablated_ Integration Ablated |
| --- | --- | --- | --- | --- | --- | --- | --- | --- | --- | --- |
| DC1 | 0.418 | 0.483 | 0.097 | 0.322 | 0.081 | 0.013 | 0.069 | 0.412 | 0.951 | 0.483 |
| DC2 | 0.859 | 0.068 | 0.014 | 0.129 | 0.105 | 0.002 | 0.069 | 0.341 | 0.905 | 0.267 |
| DC3 | 0.413 | 0.729 | 0.631 | 0.003 | 0.727 | 0.827 | 0.041 | 0.883 | 0.005 | 0.024 |
| Compelling | 0.292 | 0.853 | 0.111 | 0.012 | 0.481 | 0.321 | 0.0987 | 0.096 | 0.005 | 0.412 |
| Rich Experience | 0.121 | 0.505 | 0.067 | 0.652 | 0.251 | 0.001 | 0.028 | 0.012 | 0.298 | 0.166 |